\documentclass[twocolumn,preprintnumbers,amsmath,amssymb,article]{revtex4}

\usepackage{epsf}
\usepackage{rotating}
\usepackage{amsmath}
\usepackage{graphicx,epsfig}

\usepackage{dcolumn}
\usepackage{bm}
\usepackage{color}

\def\cD{{\cal D}}

\def\cN{{\cal N}}

\def\cH{{\cal H}}

\def\cB{{\cal B}}

\def\atd{\avg{D}}
\def\atc{\avg{C}}
\def\aS{d}
\def\attc{\avg{\tilde C}}
\def\catd{\avg{\Delta D}}
\def\catp{\avg{\Delta P}}
\def\catc{\avg{\Delta C}}

\def\catdp{\avg{\Delta D}/\avg{D}}
\def\catpp{\avg{\Delta P}/\avg{D}}
\def\catcp{\avg{\Delta C}/\avg{C}}

\newcommand{\snij}{\sigma^\nu_{ij}}
\newcommand{\snji}{\sigma^\nu_{ji}}
\newcommand{\dsnij}{\delta\sigma^\nu_{ij}}

\newcommand{\snijs}{\sigma^{\nu *}_{ij}}

\newcommand{\vs}{{\boldsymbol\sigma}}
\newcommand{\vsss}{{\boldsymbol\sigma}^{*}}
\newcommand{\dvs}{\delta{\boldsymbol\sigma}}
\newcommand{\dvss}{\delta{\boldsymbol\sigma}^{*}}

\newcommand{\vis}{{\boldsymbol I}^{*}}

\newcommand{\chd}{{\tilde\cH}}
\newcommand{\tE}{{\tilde E}}
\newcommand{\dI}{{\Delta I}}

\newcommand{\dvIs}{\delta{\boldsymbol I}^{*}}

\newcommand{\vIs}{{\boldsymbol I}^{*}}

\newcommand{\req}[1]{Eq.~(\ref{#1})}
\newcommand{\avg}[1]{\langle #1\rangle}

\newcommand{\fig}[1]{Fig.~\ref{#1}}

\DeclareMathOperator*{\argmin}{{\rm argmin}}

\newcommand{\cut}[1]{{}}
\newcommand{\etal}[1]{\emph{~et al.}}

\newcommand{\sfb}{\Omega}

\newcommand{\Lamb}{\Lambda}

\begin{document}

\preprint{}

\title[Title]
{Optimally coordinated traffic diversion by statistical physics}

\author{Tak Shing Tai and Chi Ho Yeung}
\email{chyeung@eduhk.hk}
\affiliation{Department of Science and Environmental Studies, The Education University of Hong Kong, Tai Po, Hong Kong}

\date{\today}

\begin{abstract}
Road accidents or maintenance often lead to the blockage of roads, causing severe traffic congestion. Diverted routes after road blockage are often decided individually and have no coordination. Here, we employ the cavity approach in statistical physics to obtain both analytical results and optimization algorithms to optimally divert and coordinate individual vehicle routes after road blockage. Depending on the number and the location of the blocked roads, we found that there can be a significant change in traveling path of individual vehicles, and a large increase in the average traveling distance and cost. Interestingly, traveling distance decreases but traveling cost increases for some instances of diverted traffic. By comparing networks with different topology and connectivity, we observe that the number of alternative routes play a crucial role in suppressing the increase in traveling cost after road blockage. We tested our algorithm using the England highway network and found that coordinated diversion can suppress the increase in traveling cost by as much as 66$\%$ in the scenarios studied.  These results reveal the advantages brought by the optimally coordinated traffic diversion after road blockage.
\end{abstract}


\maketitle


\section{Introduction}
Unexpected and sudden road blockage is often a result of road accidents. The accidents produce economics loss as the cars may be damaged, and the drivers and passengers may be injured \cite{CHANTITH2020}. After road accidents, the road may be blocked and  vehicles planned to travel on the road have to be diverted to other routes. The choices of diverted routes are often made individually and have no coordination. The traffic flow of the other routes may increase non-uniformly due to the lack of coordination, causing severe traffic congestion on some specific routes. This traffic congestion is often unexpected and unavoidable, since it is impossible to predict when and where road accidents and the subsequent road blockage take place.

Other than road accidents,  road maintenance works often lead to road blockage for a period of time. Although such blockage is often announced in advance, they can still cause traffic congestion. Vehicles commute in the network everyday may have to test different alternative routes and re-route several times until a new user equilibrium or Nash equilibrium of traffic is attained, which can be sub-optimal~\cite{wardrop1952, PhysRevE.103.022306}. Although such congestion after accidents or road maintenance seems to be unavoidable, an algorithm which can optimally divert traffic to suppress the congestion can help individuals save traveling cost and thus suppress economics loss. 

To examine how disruptions on links impact transportation networks, studies have been devoted on network robustness  and the identification of critical nodes and edges vulnerable for large impact. These studies examine how different attacks such as sub-optimal attack algorithms and edge-removal attacks affect network robustness  \cite{HAO2020122759,HAO2020109772,Wandelt2018,PhysRevE.98.012313}. The attacks can involve two or more steps to compare the difference between the original network  and the network with removed nodes or edges \cite{PhysRevE.86.036106,PhysRevE.92.012814}. The relationship between network structures and the robustness of the networks is also examined \cite{Wandelt2018}. Increasing network robustness is important to a wide range of networks such as  power grids, World Wide Web, transportation networks, etc., and algorithms have been proposed to enhance network robustness\cite{PhysRevE.75.036101,PhysRevE.86.066103,PhysRevE.94.022310}. Nevertheless, these studies focus on how structural properties improve network robustness, instead of remedial measures after links are removed~\cite{hu2016recovery}.

In this paper, we introduce a model to study the impact of broken links in transportation networks with initially optimized traffic flow. We apply the cavity approach originally developed for the studies of spin glasses to derive an algorithm to optimally divert the traffic after road blockage and to analytically compare various quantity of interests before and after road blockage \cite{PhysRevE.66.056126,PhysRevLett.108.208701}. Analytical results lead to good agreement with simulations. We tested the derived algorithm using the England highway network and the benefits brought by coordinated traffic diversion are observed. Our results shed light on the impact of road blockage on optimized transportation networks and the significance and the benefits brought by optimal traffic diversion.

\section{Model}
\label{sec_model}

\subsection{The original optimized networks}

We consider a model of transportation networks with $N$ nodes denoted by $i =1, \dots , N$; each node $i$ is connected with $k_i$ neighbors. There are $M$ vehicles labeled by $\nu =1, \dots, M$ on the network, and we denote the density of vehicles on the network to be $\rho=M/N$. Each vehicle $\nu$ starts from a randomly drawn origin node $O_\nu$, and travels to a common destination node $\cD$. We denote the route of vehicle $\nu$ on the link between nodes $i$ and $j$ by a variable $\snij$, such that
\begin{eqnarray}
\snij =
\begin{cases}
1, &\mbox{$\nu$ travels from $i\!\to\! j$}
\\
-1, &\mbox{$\nu$ travels from $j\!\to\!i$}
\\0, &\mbox{if $\nu$ does not travel between $i$ and $j$,}
\end{cases}
\end{eqnarray}
such that $\snji = -\snij$. The total traffic flow from node $i$ to $j$ is $|I_{ij}| = \sum_\nu |\snij|$. We further introduce the cost function 
\begin{eqnarray}
\label{eq_H}
\cH(\vs| \gamma)=\frac{1}{M}\sum_{(ij)}\Big(\sum_\nu|\snij|\Big)^\gamma=\frac{1}{M}\sum_{(ij)}|I_{ij}|^\gamma,
\end{eqnarray}
where $(ij)$ denotes the un-ordered combination of $i,j$, and with $\gamma>1$, the cost increases non-linearly with the traffic flow to discourage the sharing of a link by multiple vehicles. While our theoretical analyses can accommodate any form of cost function in terms of traffic volume $|I|$, we remark that the simple form of \req{eq_H} captures the essence in describing the time for vehicles to travel through a road with traffic congestion, which is often modeled by a polynomial function dominated by the leading power of traffic volume on the road; an example is the Bureau of Public Roads (BPR) latency function~\cite{bpr}. Nevertheless, for simplicity in the subsequent numerical analyses, we will mainly study the cases with $\gamma=1$ and $2$ as in \cite{PhysRevLett.108.208701, PhysRevE.103.022306}, which already correspond to the cases without coordination (i.e. vehicles travel via the shortest path) and with coordination respectively.

With the defined cost function $\cH$, traffic congestion can be suppressed with the path configuration 
\begin{align}
\label{eq_sigmas}
\vsss = \argmin_{\vs}\cH(\vs| \gamma)
\end{align}
which minimizes $\cH$, and can be identified by a message-passing algorithm derived in~\cite{PhysRevLett.108.208701}. With $\gamma>0$, Ref.~\cite{PhysRevLett.108.208701} also shows that vehicles traveling on the same link always head in the same direction i.e. either $\snij\ge 0$ or $\snij \le 0$ for all $\nu$ on the link $i\to j$, and the directed traffic flow between node $i$ and $j$ is given by $I_{ij}=\sum_\nu \snij$.

\subsection{Networks with broken links}

In this paper, we study a scenario in which a subset of roads fails to function, and examine the impact of the blocked roads on the coordinated traffic which optimized the cost $\cH$. Specifically, we first denote the set of disconnected links to be $\cB$, which are randomly drawn given that the remaining networks are still connected. The number of broken link is denoted as $B=|\cB|$. We then define the diversion of route for vehicle $\nu$ on the link between nodes $i$ and $j$ by a variable $\dsnij$, and examine the optimally diverted route by minimizing the cost
\begin{align}
\label{eq_H_broken}
\chd(\dvs| \vsss, \gamma)&=\frac{1}{M}\sum_{(ij)\notin \cB}\Big(\sum_\nu|\snijs + \dsnij|\Big)^\gamma
\nonumber\\
&=\frac{1}{M}\sum_{(ij)\notin \cB}|I_{ij}^* + \delta I_{ij}|^\gamma,
\end{align}
where $\vsss$ can be either the path configuration optimizing the original cost $\cH$ given by \req{eq_sigmas}, or any path configuration before road blockage, e.g. the existing traffic configuration in real transportation network which is not necessarily optimized. We will analyze $\chd$ analytically and derive a message-passing algorithm to identify the optimal diverted path configuration
\begin{align}
\dvss = \argmin\chd(\dvs| \vsss, \gamma),
\end{align}
to examine the relation between the original path configuration $\vsss$ and the diverted one $\dvs + \vsss$.

\subsection{The quantities of interest}

To characterize the impact of the diverted traffic due to the failed links on the network, we will examine the following quantities of interest both analytically and by simulations. For the original networks, we will examine the average traveling distance per vehicle, given by
\begin{align}
\label{eq_averageTravelingDistance}
\atd = \frac{1}{M} \sum_{(ij)} |I_{(ij)}^*|,
\end{align}
and the average optimized cost per vehicle, given by
\begin{align}
\label{eq_averageTravelingCost}
\atc = \frac{1}{M} \sum_{(ij)} |I_{(ij)}^*|^\gamma.
\end{align}
For networks with failed links, we will examine the changes in the above quantities. For cases with broken links on the major routes to the destination, there may be a large change in the traveling path for some vehicles but only a small change in their traveling distance, e.g. they may have to switch to another major functioning routes with similar path length. To differentiate the changes in path and distance, we define the \emph{average change in traveling path per vehicle per broken link} to be $\catp$, given by
\begin{align}
\label{eq_catp}
\catp =  \frac{1}{MB}\sum_{(ij)} |\dI_{ij}|
\end{align}
where $\dI_{ij}$ corresponds to the change of traffic flow on link $(ij)$ in the network with broken links. The \emph{average change in traveling distance per vehicle per broken link} to be $\catd$, given by
\begin{align}
\label{eq_catd}
\catd =\frac{1}{MB} \sum_{(ij)} (|I^*_{ij} + \dI_{ij}|) - \atd.
\end{align}
We note that $\catp\ge 0$ and $\catp \ge\catd$ where $|\Delta I| \ge |I+\Delta I| - |I|$; the quantity $\catd$ can be negative as we will see later in Sec. \ref{sec_multiB}. 
On the other hand, the \emph{average change of traveling cost per vehicle per broken link} is denoted as $\catc$, given by
\begin{align}
\label{eq_averageChangeOfTravelingCost}
\catc =  \frac{1}{MB} \sum_{(ij)} |I^*_{ij} + \dI_{ij}|^\gamma - \atc .
\end{align}

\section{Optimization Algorithms by the Cavity Approach}
\label{sec_algorithm}

\subsection{Algorithm for optimizing the original flows}
\label{sec_opt_flow}

To facilitate the analysis on the model, we follow the framework developed in \cite{PhysRevLett.108.208701} to map the route optimization problem to a problem of resource allocation. We first assign a load $\Lamb_i$ to each node $i$ of the network such that
\begin{align}
\label{eq_lambda}
\Lamb_i =
\begin{cases}
1, &\mbox{if $i = O_\nu, \exists\nu$;}
\\
-\infty, &\mbox{if $i = \cD$;}
\\
0, &\mbox{otherwise.}
\end{cases}
\end{align}
To ensure a single path is identified for each vehicle from its origin to the common destination, we denote the net resources of each node $i$ to be $R_i$ given by
\begin{eqnarray}
R_i = \Lambda_i + \sum_{j\in \cN_i} I_{ji},
\end{eqnarray}
where $\cN_i$ denotes the set of neighbors of node $i$. We further assume that the net resources are conserved on each node except the destination, i.e. $R_i= 0$, $\forall i \neq \cD$.

We then employ the cavity approach~\cite{mezard87} and assume that only large loops exist in the networks studied, such that neighbors of node $i$ become statistically independent if it is being removed. At zero temperature, we express the optimized energy $E_{i\to l}(I_{il})$ of a tree network terminated at node $i$ as a function of the traffic flow $I_{il}$ from $i$ to $l$. Next, we write down a recursion relation to relate $E_{i\to l}(I_{il})$ to the energy $E_{j\to i}(I_{ji})$ of its neighbors $j$ other than $l$~\cite{PhysRevLett.108.208701}, given by
\begin{align}
\label{eq_prl_recur}
E_{i\to l}(I_{il}) = \min_{\left\{ \left\{I_{ji}\right\}\left| R_i=0\right\}\right.}
\Bigg[ |I_{il}|^\gamma\!+\!\sum_{j\in\cN_i\backslash l}E_{j\to i}(I_{ji}) \Bigg].
\end{align}
We further note that $E_{i\to l}(I_{il})$ is an extensive quantity of which the value of energy is dependent on the number of nodes in the network; we therefore define an intensive quantity $E^V_{i\to l}(I_{il})$ as
\begin{align}
\label{eq_prl_EV}
E^V_{i\to l}(I_{il}) = E_{i\to l}(I_{il}) - E_{i\to l}(0).
\end{align}
Since $E^V_{i\to l}(I_{il})$ and $E_{i\to l}(I_{il})$ only differ by a constant, in \req{eq_prl_recur}, one can use $E^V_{j\to i}(I_{ji})$ instead of $E_{j\to i}(I_{ji})$ to compute $E_{i\to l}(I_{il})$, and then use \req{eq_prl_EV} to obtain the same $E^V_{i\to l}(I_{il})$.

The above equations can be used as an optimization algorithm to identify the optimal configuration of traffic flow by a group of $M$ vehicles between their origin and a common destination. In this case, the energy function $E^V_{i\to l}(I_{il})$ is considered to be a message passed from node $i$ to $l$; the message of $E^V$ is then updated according to Eqs (\ref{eq_prl_recur}) and (\ref{eq_prl_EV}) on all links until convergence. The optimal flow $I_{il}^*$ from node $i$ to $l$ can be found by 
\begin{align}
\label{eq_prl_istar}
I_{il}^* = \argmin_I[E^V_{i\to l}(I) + E^V_{l\to i}(-I) - |I|^\gamma].
\end{align}
The optimal flow on each individual link can be identified by \req{eq_prl_istar}, which constitutes the optimal route configuration on the whole network.

\subsection{Algorithms for optimal traffic diversion}
\label{sec_opt_diversion}

To analyze how traffic are optimally diverted, we assume that the original traffic configuration is given by $\vis$, before a set $\cB$ of links become disconnected. In this case, we define $\tE_{i\to l}(\dI_{il})$ to be the optimized energy of a tree network terminated at node $i$, as a function of the change of traffic flow $\dI_{il}$ from $i$ to $l$. For functioning links $(il)\notin \cB$, we can then write down a recursion relation to relate $\tE_{i\to l}(\dI_{il})$ to the energy $\tE_{j\to i}(\dI_{ji})$ of its neighbors $j$ other than $l$, given by
\begin{align}
\label{eq_tE_recur}
&\tE_{i\to l}(\dI_{il}) 
\\
&\quad = \min_{\left\{ \left\{\dI_{ji}\right\}\left| \tilde R_i=0\right\}\right.}
\Bigg[ |I_{il}^* + \dI_{il}|^\gamma\!+\!\sum_{j\in\cN_i\backslash l}\tE_{j\to i}(\dI_{ji}) \Bigg].
\nonumber
\end{align}
Since $\tE_{i\to l}$ is an extensive quantity, we define an intensive quantity $\tE^V_{i\to l}$ given by
\begin{align}
\label{eq_tE_EV}
\tE^V_{i\to l}(\dI_{il}) = \tE_{i\to l}(\dI_{il}) - \tE_{i\to l}(0)
\end{align}
On the other hand, for broken links $(il)\in \cB$, the flows on them become zero such that the intensive $\tE^V_{i\to l}$ is given by
\begin{align}
\label{eq_tE_recur2}
\tE^V_{i\to l}(\dI_{il}) =
\begin{cases}
0, &\mbox{if $\dI_{il} = -I_{il}^*$,}
\\
\infty, &\mbox{otherwise.}
\end{cases}
\end{align}
such that $\dI_{il} = -I_{il}^*$ always.

Similar to identifying the optimal flows in the original network in Sec.~\ref{sec_opt_flow}, one can use Eqs.~(\ref{eq_tE_recur}) to (\ref{eq_tE_recur2}) to identify the optimal diverted traffic in the network with failed links. In this case, the energy function (or messages) $\tE^V$ are passed on the network and updated according to Eqs.~(\ref{eq_tE_recur}) and (\ref{eq_tE_EV}) until convergence. The optimal change of flow on link $(il)$ is given by
\begin{align}
\label{eq_prl_istar}
\dI_{il}^* = \argmin_\dI[\tE^V_{i\to l}(\dI) + \tE^V_{l\to i}(-\dI) - |\dI|^\gamma].
\end{align}
According to \req{eq_tE_recur2}, for broken link $(il)\in\cB$, \req{eq_prl_istar} gives $\dI_{il}^* = -I_{il}^*$. The optimal diverted traffic $\dvIs$ for each individual link and thus the whole network can be identified by \req{eq_prl_istar}, the new configuration of flows on the network with broken links is given by $\vIs + \dvIs$

Indeed, the original traffic optimization algorithm \req{eq_prl_recur} can be used to obtain the new traffic configuration in the network with broken links, but our newly proposed algorithm \req{eq_tE_recur} has the following advantages over \req{eq_prl_recur}: (1) traffic can be optimally diverted subject to an objective function in terms of the changes in traffic flow in \req{eq_tE_recur}, for instance, minimizing the changes via $\sum_{(ij)}|dI_{ij}|$; (2) \req{eq_tE_recur} can be used to divert any initial traffic configuration $\{I^*_{ij}\}$, even if $\{I^*_{ij}\}$ is not optimized; (3) the computational cost of \req{eq_tE_recur} is much lower than that of \req{eq_prl_recur} as we will see in Sec.~\ref{sec_computation}.

\subsection{Reduction of computation cost}
For networks with broken links,  we denote $\Omega$ to be the sum of the magnitude of traffic flow on all broken links in the initial network, given by 
\begin{align}
\label{eq_range}
\Omega =\sum_{(ij)\in \cB}^{} |I^*_{ij}|
\end{align}
In cases with $\gamma>1$,  $\dI_{ij}$ is limited to the range $[-\Omega, \Omega]$, while $I_{ij}$ is limited to the range $[-M,M]$. Since  $\Omega\ll M$ when the number of broken links is small, for instance $B=1$, the computation cost of the traffic diversion algorithm \req{eq_tE_recur} is much lower than that of the original algorithm \req{eq_prl_recur} when it is used on the network with broken links. In addition, \req{eq_tE_recur} can be used for networks given any initial configuration of flow $\{I^*_{ij}\}$, even if this initial configuration is not optimized.

\section{Analytical solutions}

\subsection{The original optimized networks}

To obtain the analytical results of the original optimized network, we have to solve for the functional probability distribution $P[E^V(I)]$. By using the recursion relation in~\req{eq_prl_recur}, we can write a self-consistent equation in terms of $P[E^V(I)]$, by summing over all degrees $k$, in the form of
\begin{align}
\label{eq_prl_prob}
&P[E^V(I)]\!=\!\sum_{k=1}^{\infty} \frac{P(k) k}{\avg{k}} \int d\Lambda P(\Lambda) \prod_{j=1}^{k-1} \int dE_j^V P[E_j^V(I)] 
\nonumber\\
&\times\delta\Bigg(E^V(I)-\min_{\left\{ \left\{I_{j}\right\}\left| \Lambda -I + \sum_j I_j =0\right\}\right.}
\Bigg[ |I|^\gamma\!+\!\sum_{j=1}^{k-1}E_{j}(I_{j}) \Bigg]\Bigg).
\end{align}
where $P(k)$ and $\avg{k}$ represent the degree distribution and its average, respectively.

One can solve \req{eq_prl_prob} numerically to obtain a stable $P[E^V(I)]$ and to compute various quantities of interests. For instance, we can compute the distribution $P(I)$ of flow $I$ on a link in the optimized network via
\begin{align}
P(I) &= \int dE_1^V P[E_1^V(I)] \int dE_2^V P[E_2^V(I)] 
\nonumber\\
&\times\delta\Big(I - \argmin_I[E^V_1(I) + E^V_2(-I) - |I|^\gamma]\Big).
\end{align}
The average traveling distance per vehicle $\atd$ and the average optimized cost $\atc$ are then given by 
\begin{align}
\atd = \frac{1}{M}\sum_I |I| P(I),
\\
\atc = \frac{1}{M}\sum_I |I|^\gamma P(I).
\end{align}
Alternatively, one can compute the increase in the optimal cost for an additional node and an additional link to the network, respectively given by
\begin{align}
\label{eq_cnode}
\avg{C_{\rm node}} =& \sum_{k=1}^{\infty} P(k) \int d\Lambda P(\Lambda) \prod_{j=1}^{k} \int dE_j^V P[E_j^V(I_j)] 
\nonumber\\
&\quad\times\min_{\left\{ \left\{I_{j}\right\}\left| \Lambda + \sum_j I_j =0\right\}\right.}
\Bigg[ \sum_{j=1}^{k}E_{j}(I_{j}) \Bigg],
\\
\label{eq_clink}
\avg{C_{\rm link}} =& \int dE_1^V P[E_1^V(I)] \int dE_2^V P[E_2^V(I)] 
\nonumber\\
&\quad\times\min_I[E^V_1(I) + E^V_2(-I) - |I|^\gamma],
\end{align}
such that $\atc = \avg{C_{\rm node}} - \frac{\avg{k}}{2}\avg{C_{\rm link}}$~\cite{PhysRevE.66.056126}. The results obtained by numerically solving the above equations can be found in \cite{PhysRevLett.108.208701}.

\subsection{The networks with broken links}
\label{sec_cavity_broken}

To analytically reveal the changes in various quantities of interests between the network with and without broken links,  we obtain the joint distribution $P[E^V(I), \tE^V(I')]$ by using the recursion relation in \req{eq_tE_recur} to write a self-consistent equation in terms of $P[E^V(I), \tE^V(I')]$, given by
\begin{align}
\label{eq_prl_recur_b}
&P[E^V(I), \tE^V(I')]= \sum_{k=1}^{\infty} \frac{P(k) k}{\avg{k}} \int d\Lamb P(\Lamb) 
\nonumber\\
&\times \prod_{j=1}^{k-1}\int dE^V(I) d\tE^V(I') P[E_j^V(I),\tE_j^V(I')]
\nonumber\\
&\times \prod_{j=1}^{k-1} \sum_{a_j=0,1} \bigg[\left(\frac{B}{Nk/2}\right) \delta_{a_j,0}+\left(1-\frac{B}{Nk/2}\right) \delta_{a_j,1}\bigg]
\nonumber\\
&\times\delta\Bigg(E^V(I) -\min_{\left\{ \left\{I_{j}\right\}\left| \Lambda -I + \sum_j I_j =0\right\}\right.}
\Bigg[ |I|^\gamma\!+\!\sum_{j=1}^{k-1}E_{j}(I_{j}) \Bigg]
\Bigg).
\nonumber\\
&\times\delta\Bigg(\tE^V(I') -\min_{\left\{ \left\{I_{j}\right\}\left| \Lambda -I' + \sum_j a_j I'_j =0\right\}\right.}
\Bigg[ |I'|^\gamma\!+\!\sum_{j=1}^{k-1}a_j \tE_{j}(I'_{j}) \Bigg]
\Bigg).
\end{align}
where $a_j=0, 1$ denotes a broken and a functioning link respectively to neighbor $j$ in the network. One can solve \req{eq_prl_recur_b} numerically to obtain a stable $P[E^V(I), \tE^V(I')]$ and to compute the joint distribution $P(I,I')$ of flow $I$ on a link in the original network and the new flow $I'$ on the same link in the network with broken links, given by
\begin{align}
P(I,I') = &\int dE_1^V d\tE_1^V P[E_1^V(I),\tE_1^V(I')]
\nonumber\\
&\times\int dE_2^V d\tE_2^V P[E_2^V(I), \tE_2^V(I')] 
\nonumber\\
&\times\delta\Big(I - \argmin_I[E^V_1(I) + E^V_2(-I) - |I|^\gamma]\Big)
\nonumber\\
&\times \left(1-\frac{B}{Nk/2}\right) \times \delta\Bigg[\Big(I' - \argmin_{I'}[\tE^V_1(I')
\nonumber\\
& + \tE^V_2(-I') - |I'|^\gamma]\Big)\Bigg] 
\nonumber\\
&\times \delta \Big(\frac{B}{Nk/2}\Big)\times\delta \Big(I'-0\Big).
\end{align}
To obtain various quantities of interests for the network with broken links, we marginalize $ P(I,I')$ and obtain $P(I')$, i.e. the distribution of optimal traffic flow in the network with broken links, given by
\begin{align}
P(I') = \sum_{I}P(I,I').
\end{align}

The average traveling distance $\atd$ and the average optimized cost $\atc$ per vehicle in the network with broken links are given by 
\begin{align}
\avg{\tilde D} = \frac{1}{M}\sum_{I'} |I'| P(I'),
\\
\avg{\tilde C} = \frac{1}{M}\sum_{I'} |I'|^\gamma P(I').
\end{align}

The average change in traveling path per vehicle per broken link, i.e. $\catp$, is given by
\begin{align}
\avg{\Delta P} = \frac{1}{MB}\sum_{I,I'}|I-I'|P(I,I'),
\end{align}

Then, the average change in traveling distance per vehicle per broken link, i.e. $\catd$, is given by
\begin{align}
\avg{\Delta D} = \frac{1}{MB}\sum_{I,I'}(|I|-|I'|)P(I,I')=\frac{1}{B}(\avg{\tilde D} - \avg{D}),
\end{align}

Alternatively, similar to Eqs.~(\ref{eq_cnode}) and (\ref{eq_clink}), one can compute $\avg{\tilde C}$ by computing $\avg{\tilde C_{\rm node}}$ and $\avg{\tilde C_{\rm link}}$, i.e. the increase in the optimal cost for an additional node and an addition link in the network with broken link,  given by
\begin{align}
\avg{\tilde C_{\rm node}&} = \sum_{k=1}^{\infty} P(k) \int d\Lambda P(\Lambda)
\nonumber\\
&\quad\times \prod_{j=1}^{k} \int dE_j^V d\tE_j^V P[E_j^V(I_j), \tE_j^V(I'_j)] 
\nonumber\\
&\quad\times \prod_{j=1}^{k} \sum_{a_j=0,1} \Bigg[\Big(\frac{B}{Nk/2}\Big) \delta_{a_j,0}+\Big(1-\frac{B}{Nk/2}\Big) \delta_{a_j,1}\Bigg]
\nonumber\\
&\quad\times\min_{\left\{ \left\{I'_{j}\right\}\left| \Lambda + \sum_j a_j I'_j =0\right\}\right.}
\Bigg[ \sum_{j=1}^{k} a_j \tE_{j}(I'_{j}) \Bigg],
\\
\avg{\tilde C_{\rm link}&} = \int dE_1^V d\tE_1^V P[E_1^V(I), \tE_1^V(I')] 
\nonumber\\
&\quad\times \int dE_2^V d\tE_2^V P[E_2^V(I), \tE_2^V(I')] 
\nonumber\\
&\quad\times\min_{I'}[\tE^V_1(I') + \tE^V_2(-I') - |I'|^\gamma],
\end{align}
such that $\attc = \avg{\tilde C_{\rm node}} - (1-\frac{B}{Nk/2})(\frac{\avg{k}}{2})\avg{\tilde C_{\rm link}}$. The average change in the cost per vehicle per broken link can be obtained by $\catc = (\attc - \atc)/B$.

\section{Results}

In this section, we will show various quantities of interest such as the average change in traveling distance and cost in the networks with broken links, obtained by the theoretical analyses described in Section~\ref{sec_cavity_broken} as well as computer simulations. For computer simulations, we iterate the Eqs. (\ref{eq_prl_recur}) and (\ref{eq_tE_recur}) on real instances of networks for $5 \times 10^6$ steps and terminate the simulations when the change of $E^V_{il}$ and $\tE^V_{il}$ is less than $1 \times 10^{-8}$ for $100 \times N$ consecutive steps.

\subsection{The impact of broken links with optimal traffic diversion}
\label{sec_multiB}

\subsubsection{Traveling path}

We first examine how the extent of road failures, i.e. the number of broken links $B$, impact on the traveling path of vehicles after optimal traffic diversion. We remark that according to \req{eq_catp}, $\catp\ge 0$, such that the average percentage change of traveling path is always greater than or equal to zero, i.e. $\catpp\ge 0$. As shown in Fig. \ref{fig_changeOfTravelingDistanceAndPath}(a), both analytical and simulation results show that  $\catpp>0$, implying that broken links always change the traveling path of some vehicles. In addition, $\catpp$ have a similar value at different vehicle density $\rho$, implying that vehicles share a similar amount of changes in traveling path per broken link even if the density of vehicle is different. Furthermore, $\catpp$ decreases when $B$ increases, implying that when the number of broken links increases, the impact of each broken link becomes less significant. 

\begin{figure*}
\includegraphics[ width=0.4\linewidth] {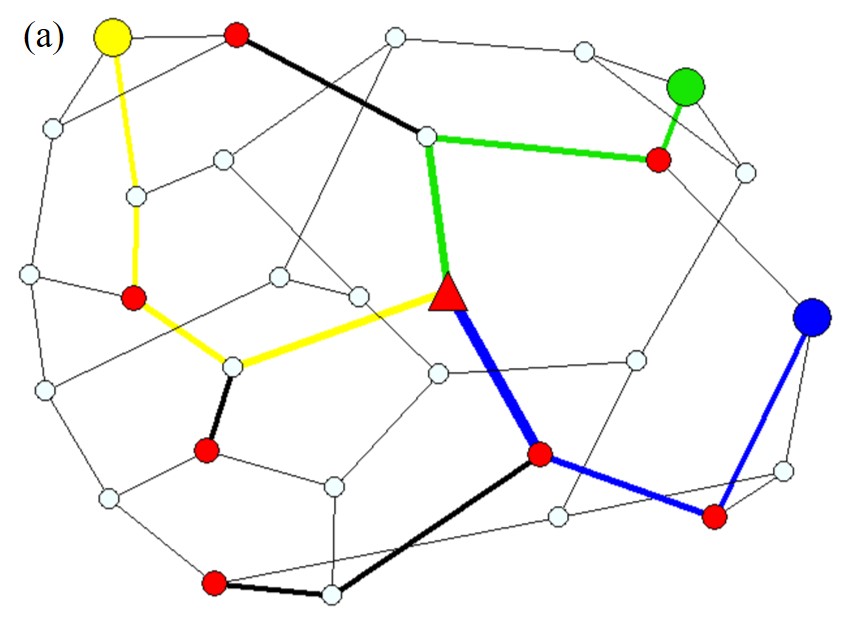}
\includegraphics[ width=0.4\linewidth] {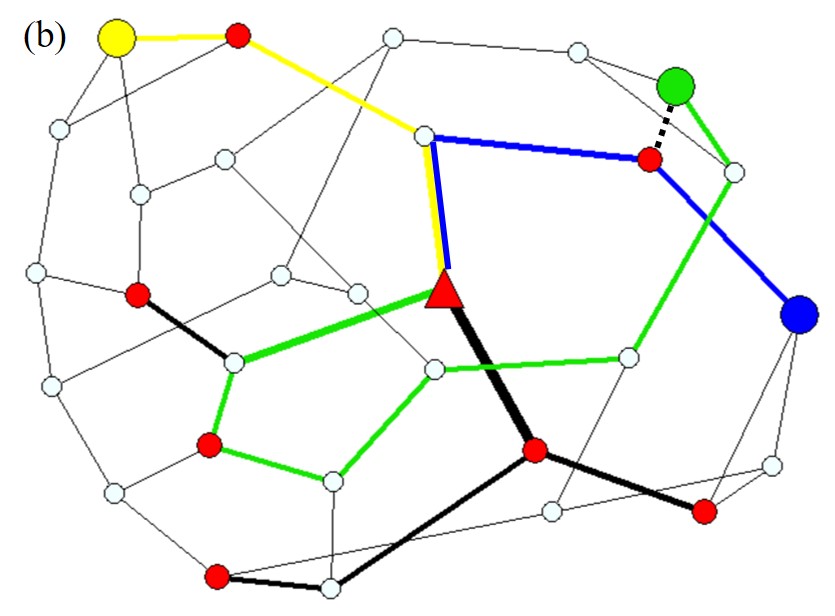}
\caption{
(a) An example of optimized path configuration for $M=10$ vehicles in a random regular graph with $N=30$ and $k=3$.  (b) The corresponding optimally diverted path configuration with one broken link, i.e. $B=1$, in the network. Filled (red) circles correspond to the origin of vehicles, and the triangle represents their common destination. The thickness of links represents the magnitude of traffic flow, and the dotted link in (b) represents the broken link. The green, yellow and blue vehicles change their paths in (b) after the link is broken.
}
\label{fig_caseOne}
\end{figure*}

Interestingly, our results show that other than the vehicles on the broken links, the traveling path of vehicles outside the broken links may have to be shifted and coordinated to achieve optimal traffic diversion. As shown in Fig. \ref{fig_caseOne}(a) and (b), when the link indicated by the dotted line in the network is broken, which is used by the green vehicle, the traveling path of two other vehicles, i.e. the blue and the yellow vehicles, have to be shifted for coordination. In this case, all traveling path, distance and cost increase. This example shows the importance of an optimization algorithm for traffic diversion, since the traveling path of all vehicles have to be coordinated to achieve optimal traffic diversion.
\subsubsection{Traveling distance}
Next, we examine the changes in the traveling distance of vehicles when there are broken links in the network. In Fig. \ref{fig_changeOfTravelingDistanceAndPath}(b), the average percentage change of traveling distance is always positive, i.e. $\catdp>0$, implying that broken links have a negative impact on the network. Moreover, we observe that the percentage change in traveling distance is smaller than that in traveling path, i.e. $\catdp > \catpp$, as expected (see explanation after \req{eq_catd} for details). We remark that $\catp$ corresponds to the changes in selected routes by individual vehicles but $\catd$ only measures the difference in the total path length by all vehicles, so $\catpp$ is different from $\catdp$. As shown in Fig. \ref{fig_changeOfTravelingDistanceAndPath}(b),  $\catdp\approx 1\%$ per broken link and is roughly independent of $B$, implying that the average path length in the network with broken links is only slightly longer than that in the original network. These results suggest that on average, vehicles are diverted to new paths with longer distance and the distance of the diverted paths increases with the number of broken links. 

\begin{figure*}
\includegraphics[ width=0.4\linewidth] {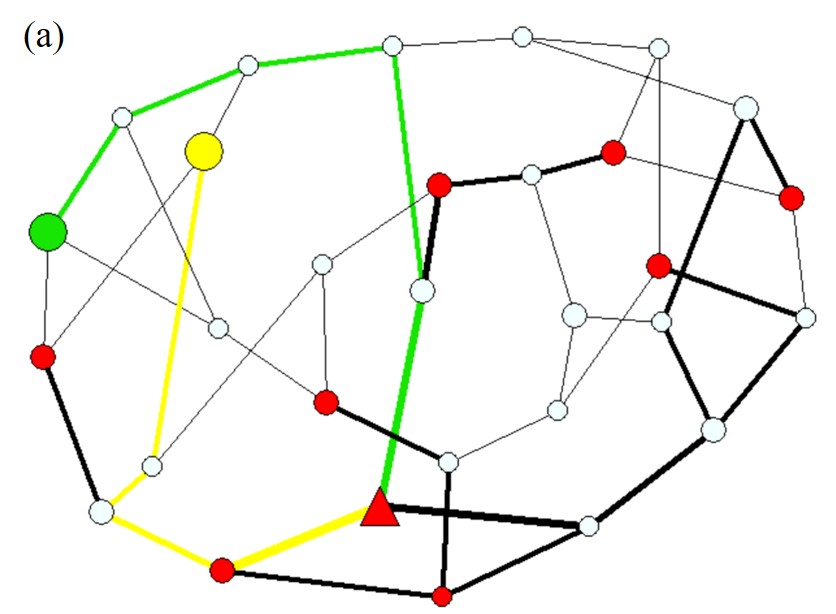}
\includegraphics[ width=0.4\linewidth] {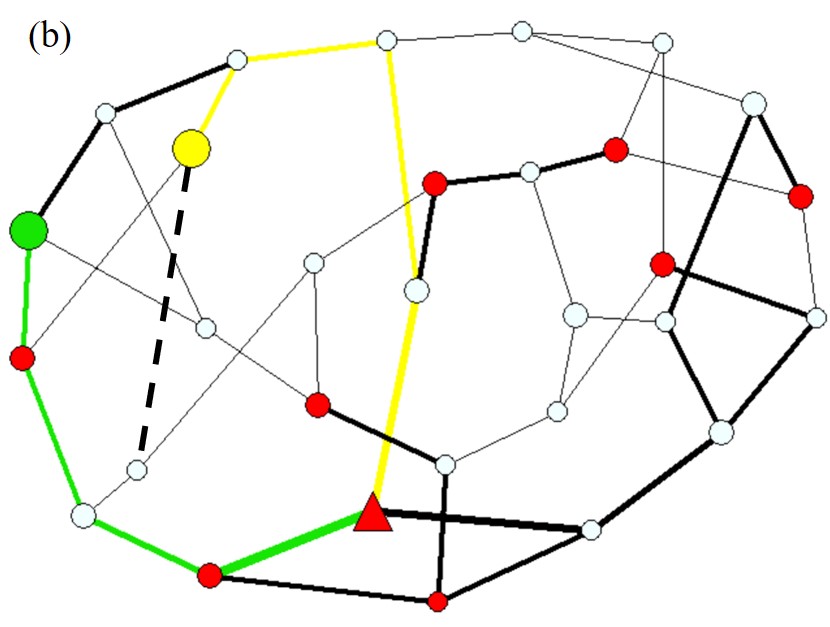}
\caption{
(b) Another example of optimized path configuration for $M=10$ vehicles in a random regular graph with $N=30$ and $k=3$. (b) The corresponding optimally diverted path configuration with one broken link, i.e. $B=1$, in the network. Filled (red) circles correspond to the origin of vehicles, and the triangle represents their common destination. The thickness of links represents the magnitude of traffic flow, and the dotted link in (b) represents the broken link. The green and yellow vehicles change their paths in (b) after the link is broken. The traveling distance of the green vehicle, and hence the average traveling distance, is shorter when there is one broken link in the network.
}
\label{fig_caseTwo}
\end{figure*}

Interestingly, there are exceptional cases where the total traveling distance decreases after some links are broken in the network. As shown in Fig. \ref{fig_caseTwo}, other than the green vehicle which initially travels on the broken link, the yellow vehicle also re-routes due to coordination. In this case,  the green vehicle is diverted to use links which are occupied by existing vehicles in the network, causing an increase in traveling cost but a decrease in traveling distance. Though the route of the yellow vehicle changes, its traveling distance does not. As a result, the overall traveling distance decreases but the traveling cost increases. We remark that this is a special case as the traveling distance increases on average as shown in \ref{fig_changeOfTravelingDistanceAndPath}(b). 

\subsubsection{Traveling cost}
\begin{figure}
\includegraphics[ width=0.9\linewidth] {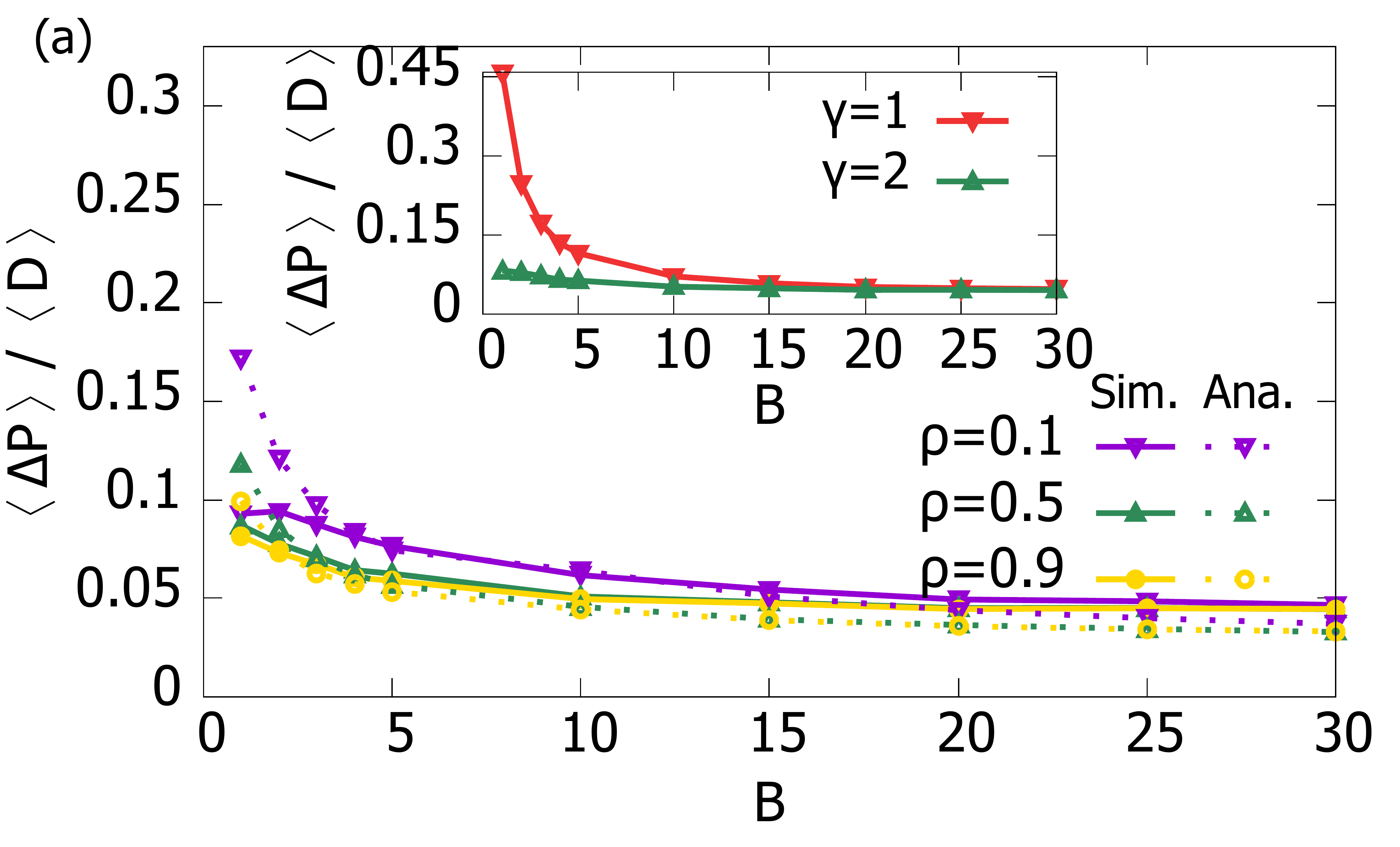}
\includegraphics[ width=0.9\linewidth] {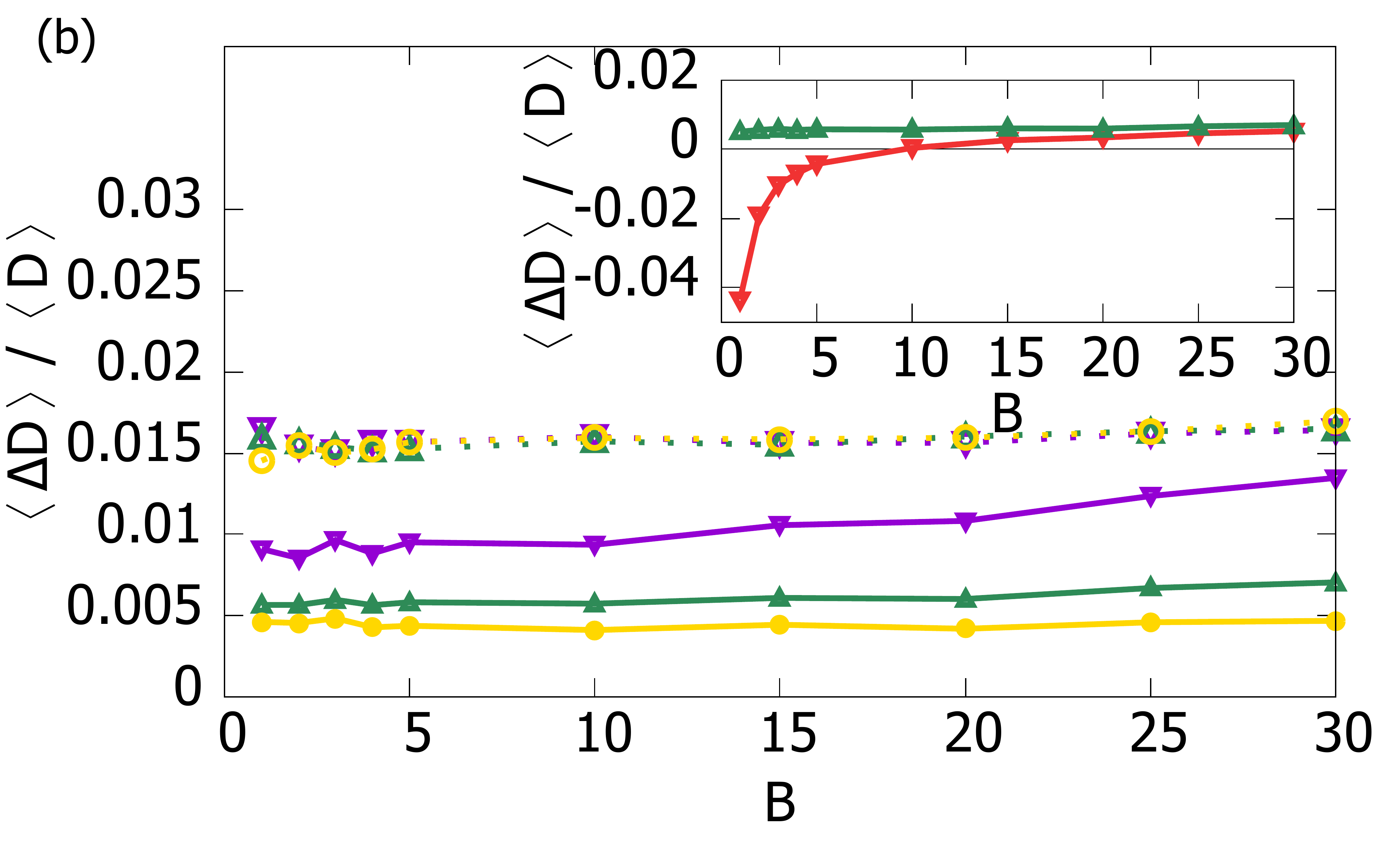}
\includegraphics[ width=0.9\linewidth] {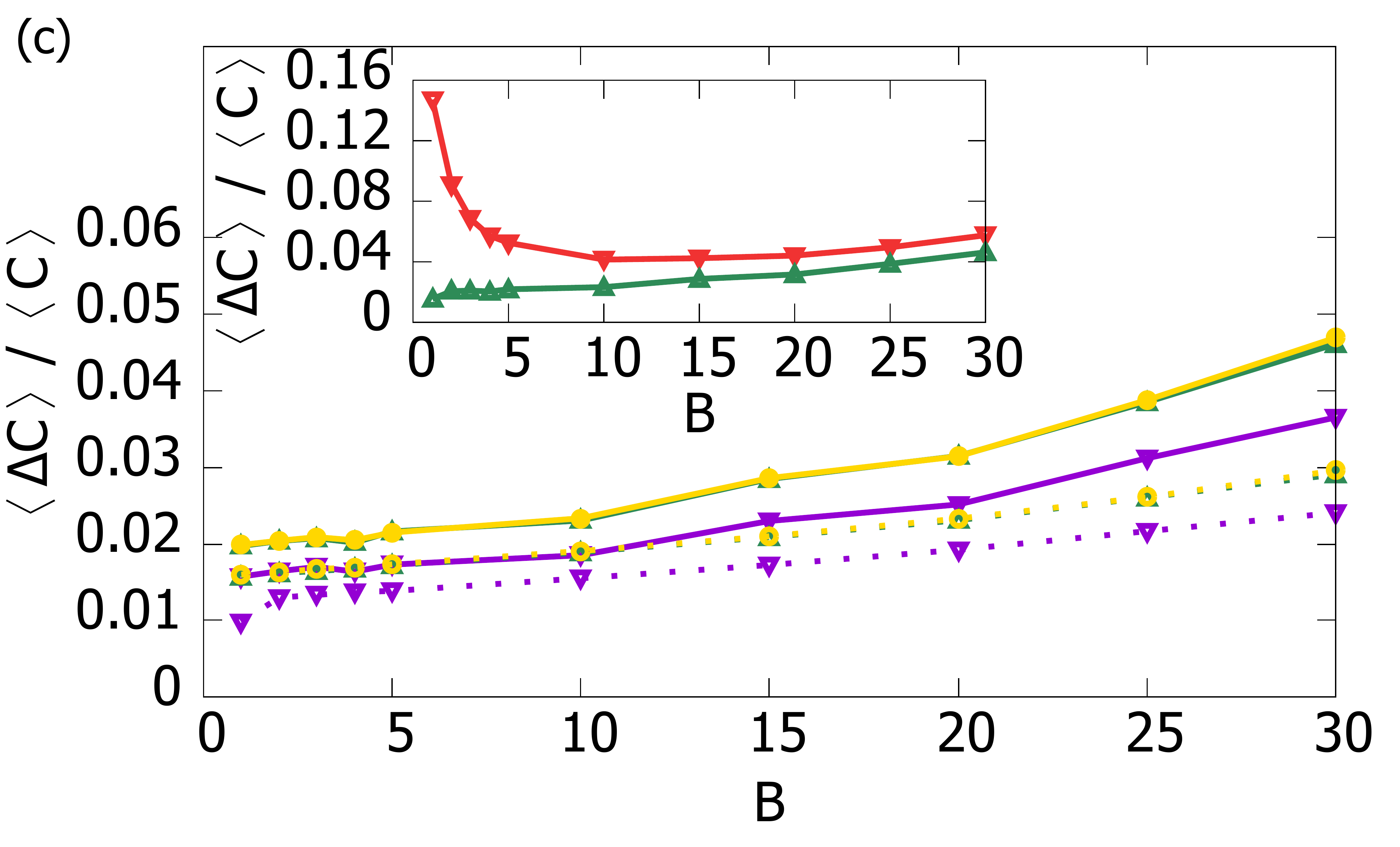}
\caption{
Analytical (dashed lines) and simulation results (solid lines) of the percentage change in (a) traveling path $\catpp$, (b) traveling distance $\catdp$, and (c) traveling cost $\catcp$, as a function of $B$, i.e. the number of broken links in network,  for various values of vehicle density $\rho$ in random regular graphs with $N=100$ nodes, $k=3$ neighbors and one common destination. The results are averaged over 1000 realizations and over 98\% samples converge within $5 \times 10^4$ updates per node.  Inset: The percentage change in (a) traveling path $\catpp$, (b) traveling distance $\catdp$, and (c) traveling cost $\catcp$, as a function of $B$ for  $\gamma =1, 2$ in the traveling cost \req{eq_H}, with $\rho=0.5$ in random regular graphs with $N=100$ nodes and $k=3$ neighbors over 1000 realizations. The percentage change in cost $\catcp$ in the cases with $\gamma=2$ is always smaller than that in the cases with $\gamma=1$, implying that coordinated traffic diversion often leads to a smaller cost increment compared to the case when vehicles are diverted to their shortest path.
}
\label{fig_changeOfTravelingDistanceAndPath}
\end{figure}

Figure \ref{fig_changeOfTravelingDistanceAndPath}(c) shows that the average percentage change in traveling cost $\catcp$ increases when vehicle density $\rho$ increases. Since links in the networks are more occupied when there are more vehicles in the network,  it is less likely for vehicles to shift to a less occupied route, hence $\catcp$ increases when $\rho$ increases. On the other hand, when $B$ increases, $\catcp$ also increases, implying that more broken links lead to a proportionately higher increase in cost. It is because when there are more broken links, fewer links can be used to divert the traffic flow and vehicles have to travel in more occupied links. Although there are discrepancies between the analytical and the simulation results, they both show a similar trend. We will further explain these discrepancies in Sec.~\ref{sec_discrepancy}.

\subsection{The dependence on the distance between the broken links and the destination}
\label{sec_distanceBtwDestinationAndB}

\begin{figure}
\includegraphics[ width=0.9\linewidth] {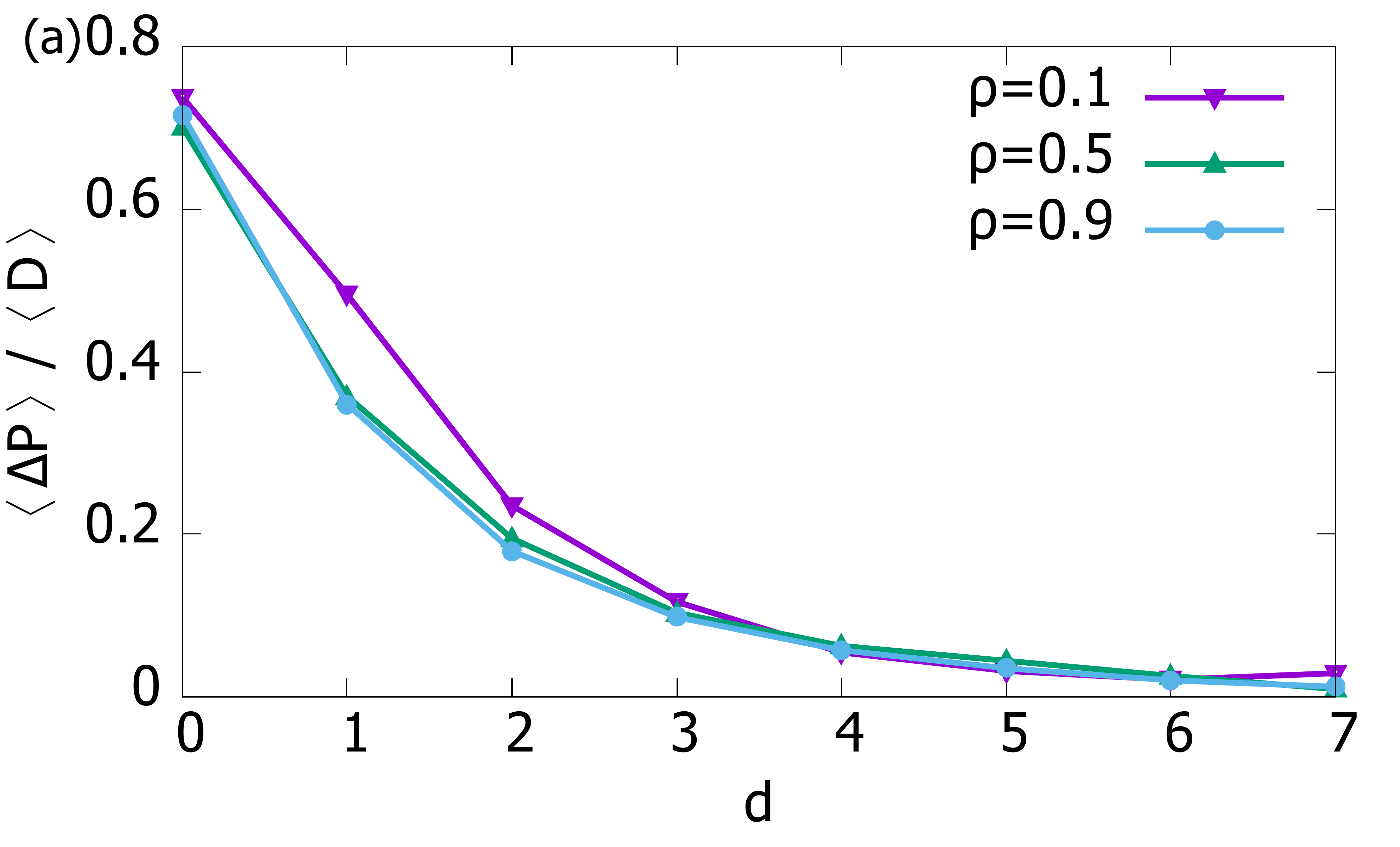}
\includegraphics[ width=0.9\linewidth] {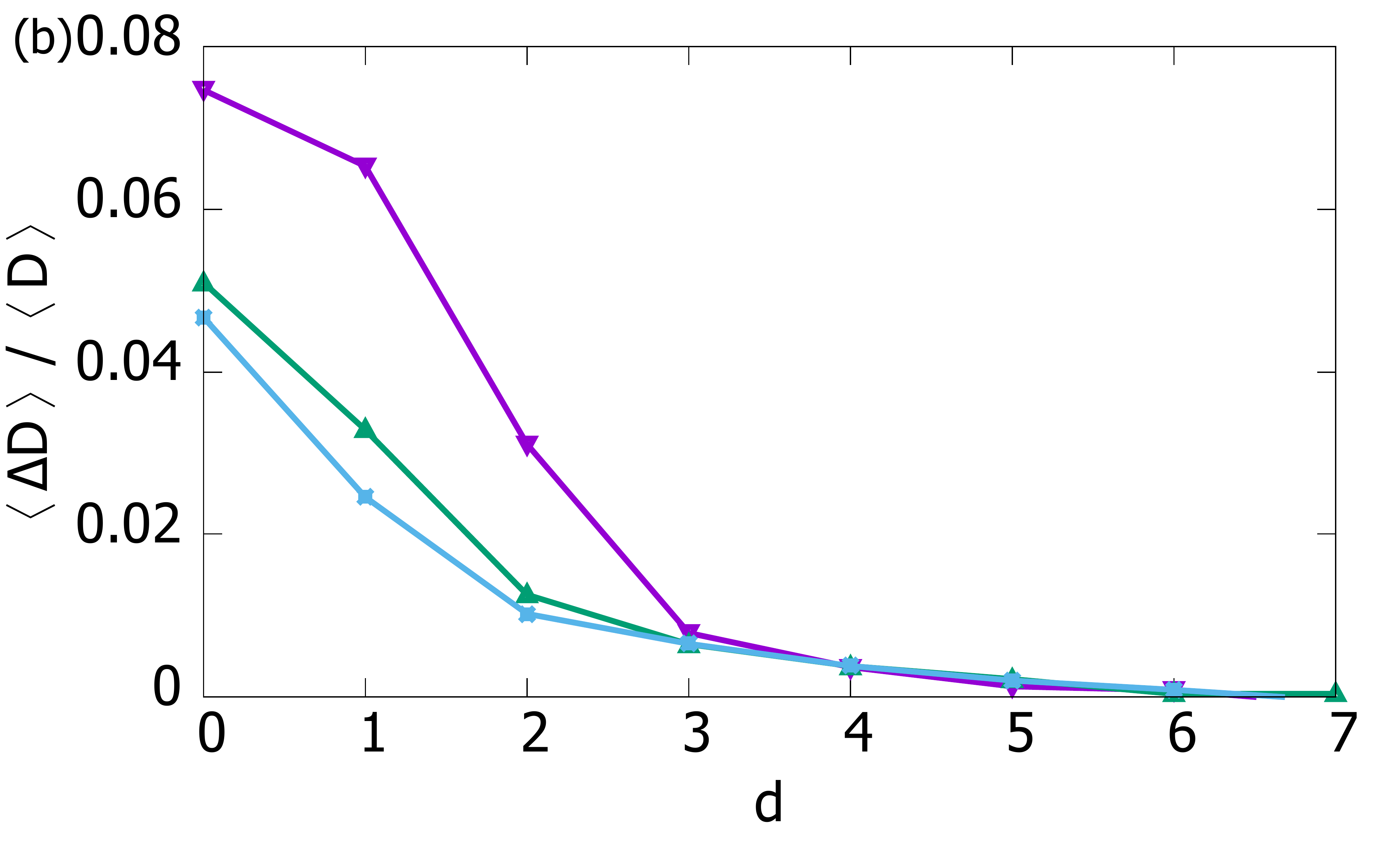}
\includegraphics[ width=0.9\linewidth] {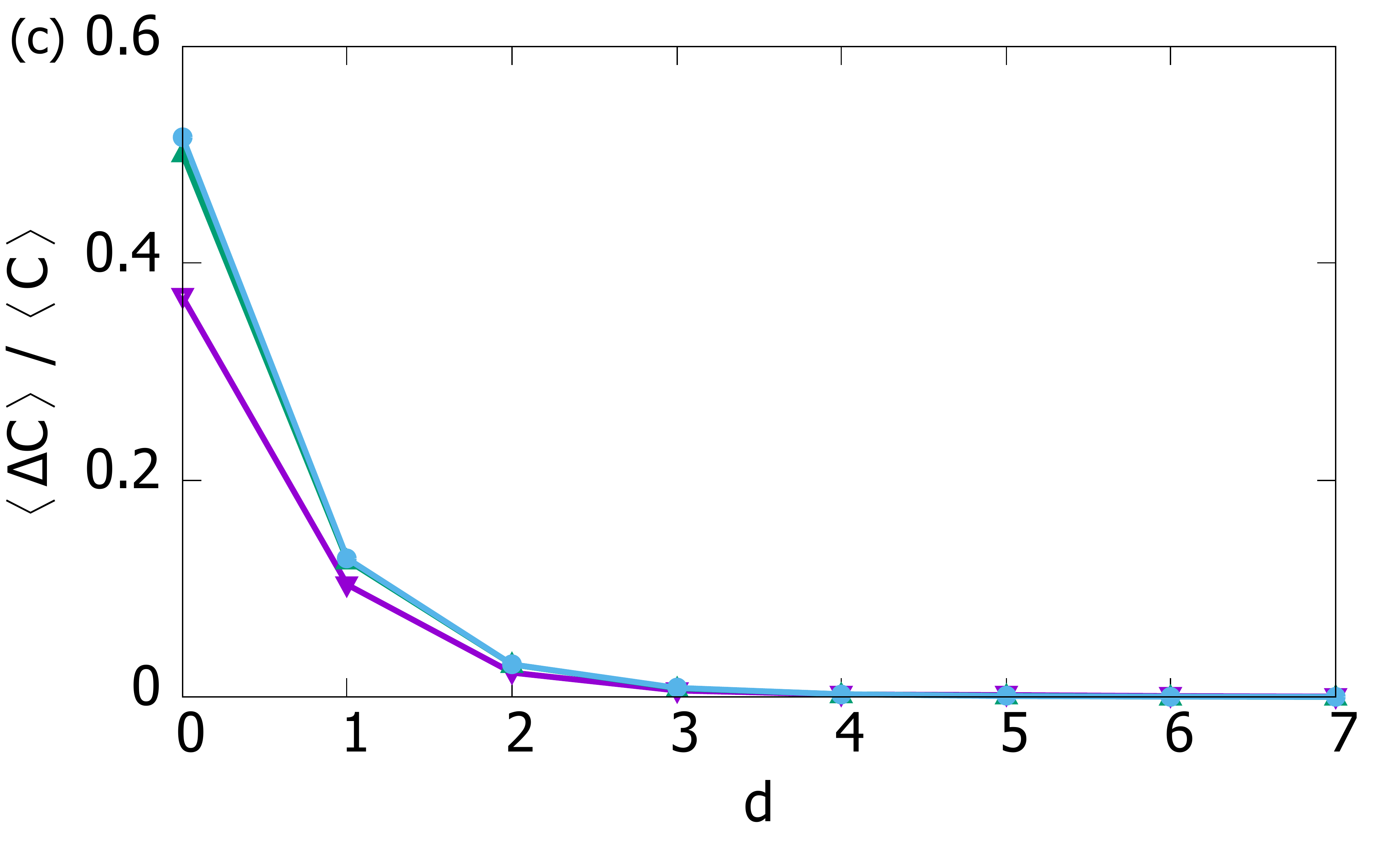}
\caption{
The percentage change in (a) traveling path $\catpp$, (b) traveling distance $\catdp$, and (c) traveling cost $\catcp$, as a function of the distance $\aS$ between the broken link and the common destination with $B=1$ for various vehicle density $\rho$. 
}
\label{fig_changeOfTravelingPathAndCostByDistance}
\end{figure}

After examining how the  number of broken links affects the optimally diverted traffic, we go on to reveal the dependence of the system on the distance $\aS$ of the broken links from the common destination. For simplicity, we set $B=1$ in the study of the impact of $\aS$. As shown in Fig. \ref{fig_changeOfTravelingPathAndCostByDistance}(a), when $\aS$ is small, $\catpp$ is large. It is because when the broken link is directly connected to the destination, a large number of vehicles need to re-route which makes $\catpp$ large. However, when $\aS$ increases, $\catpp$ decreases. These results imply that the influence of the broken link on the vehicles decreases when the broken link is far away from the destination. Moreover,  $\catpp$ at different vehicle density $\rho$ shows a similar value at the same distance $\aS$, implying $\aS$ is a more crucial factor for $\catpp$ than $\rho$. 

Similar to $\catpp$, as shown in Fig \ref{fig_changeOfTravelingPathAndCostByDistance}(b), $\catdp$ is the highest when the broken link is directly connected to the destination. When $\aS$ increases, $\catdp$ decreases. It is because when the distance between the broken link and the common destination increases, the number of possible paths to the destination increases. Therefore, when $\aS$ increases, there is a higher chance for vehicles to find alternative paths with similar distance to the destination, and $\catdp$ decreases. 

Finally, we see in Fig. \ref{fig_changeOfTravelingPathAndCostByDistance}(c) the percentage change in the traveling cost increases when$\aS$ decreases, implying that a broken link near the common destination greatly increases the traveling cost. We can also see that the percentage change of traveling cost at different vehicle density $\rho$ but the same $\aS$ is similar, again implying that $\catcp$ depends more strongly on $\aS$ instead of $\rho$. When $\aS$ increases, the percentage change of traveling cost is very small, suggesting that the traveling cost of vehicles have almost no change when the broken link is far away from the destination.

The above results show that the location of the broken links is closely related to the amount of traffic needed to be diverted. As expected, it implies that when the blocked roads are close to the city center, there is a significant change in the traveling path of individual vehicles as well as a significant increase in their average traveling distance and cost. In this case, an optimization algorithm for traffic diversion would be most valuable when a large amount of traffic have to be diverted and coordinated.

\subsection{The dependence on network topology}
\label{sec_diffk}
\begin{figure}
\includegraphics[ width=0.9\linewidth] {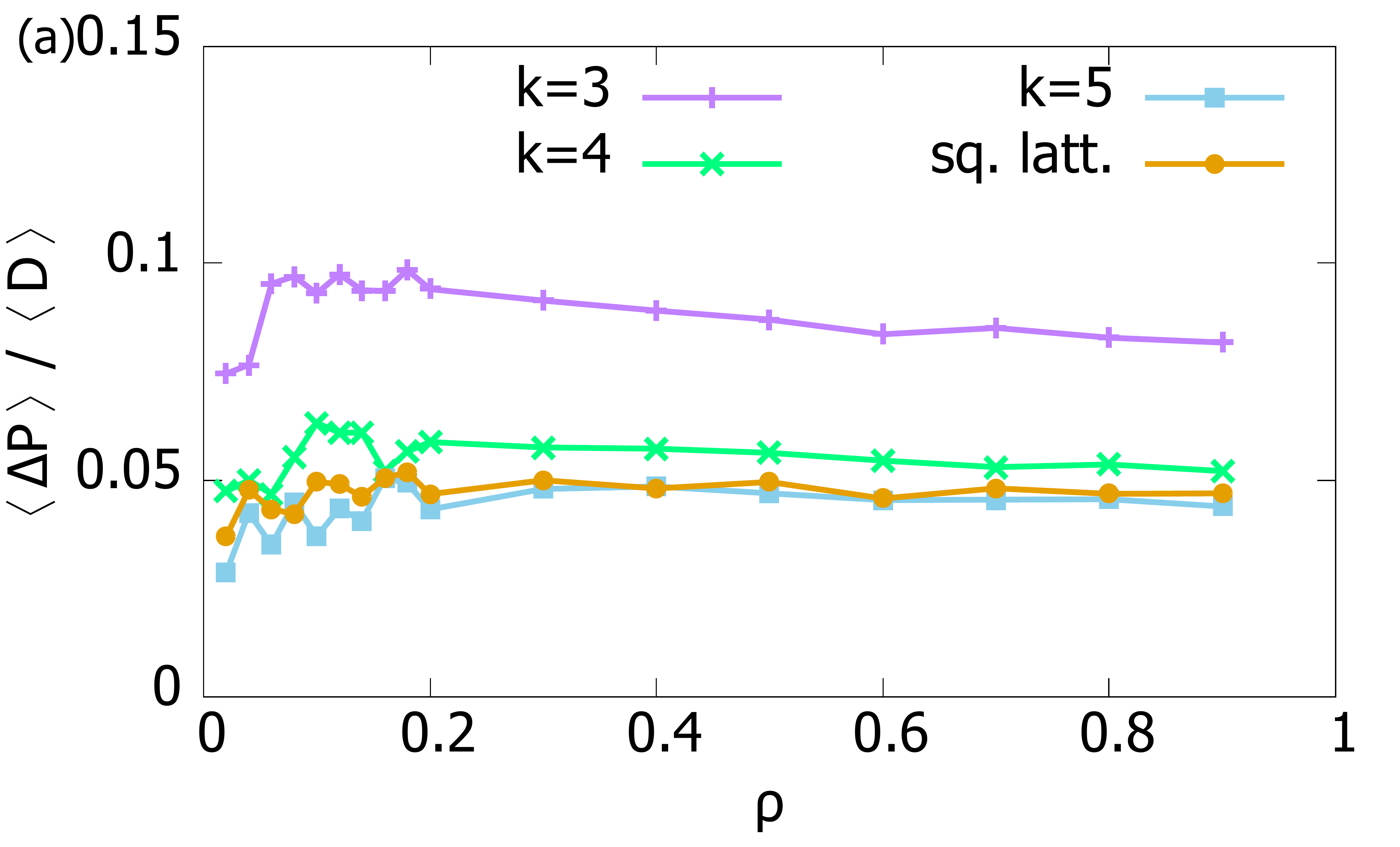}
\includegraphics[ width=0.9\linewidth] {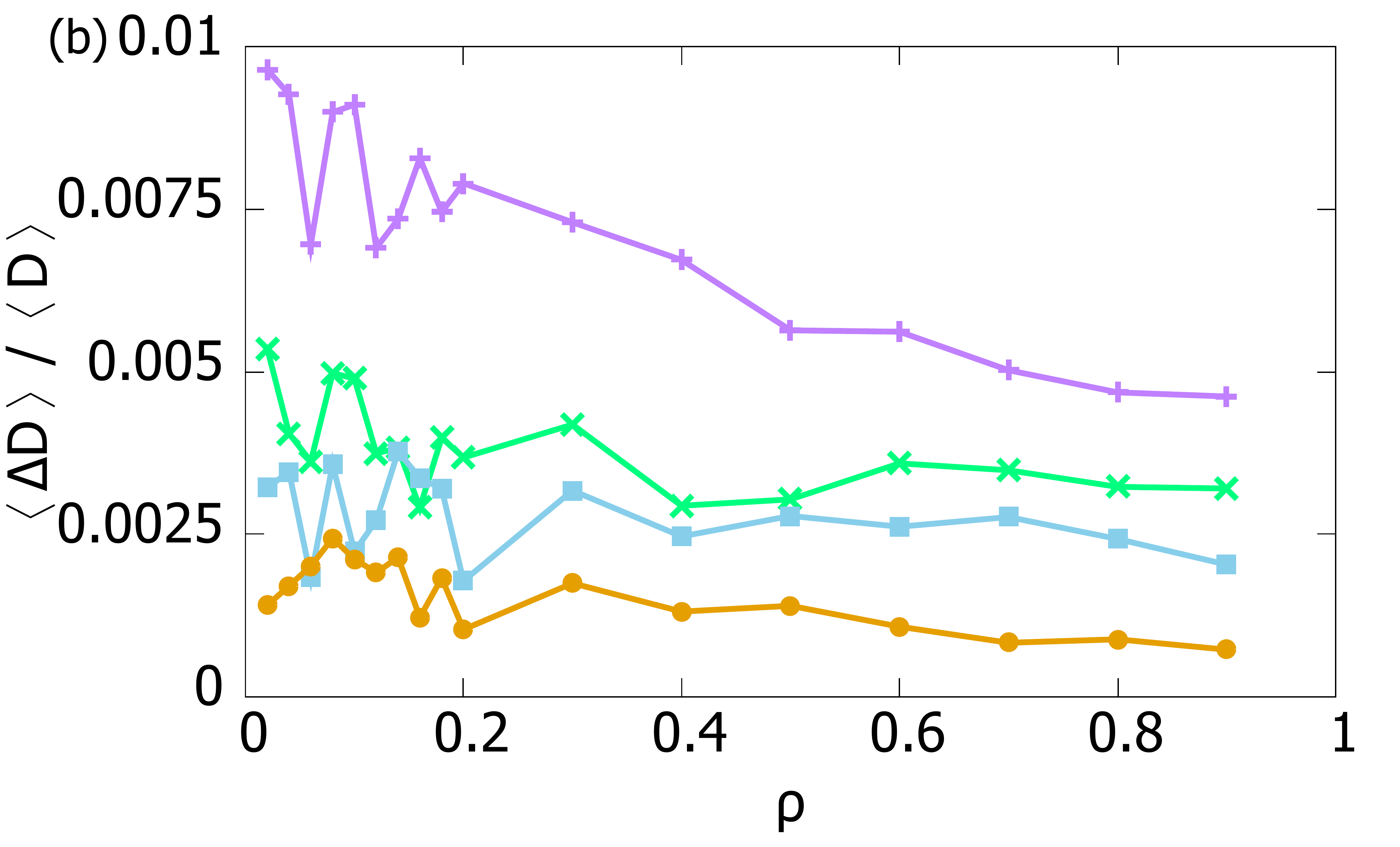}

\includegraphics[ width=0.9\linewidth] {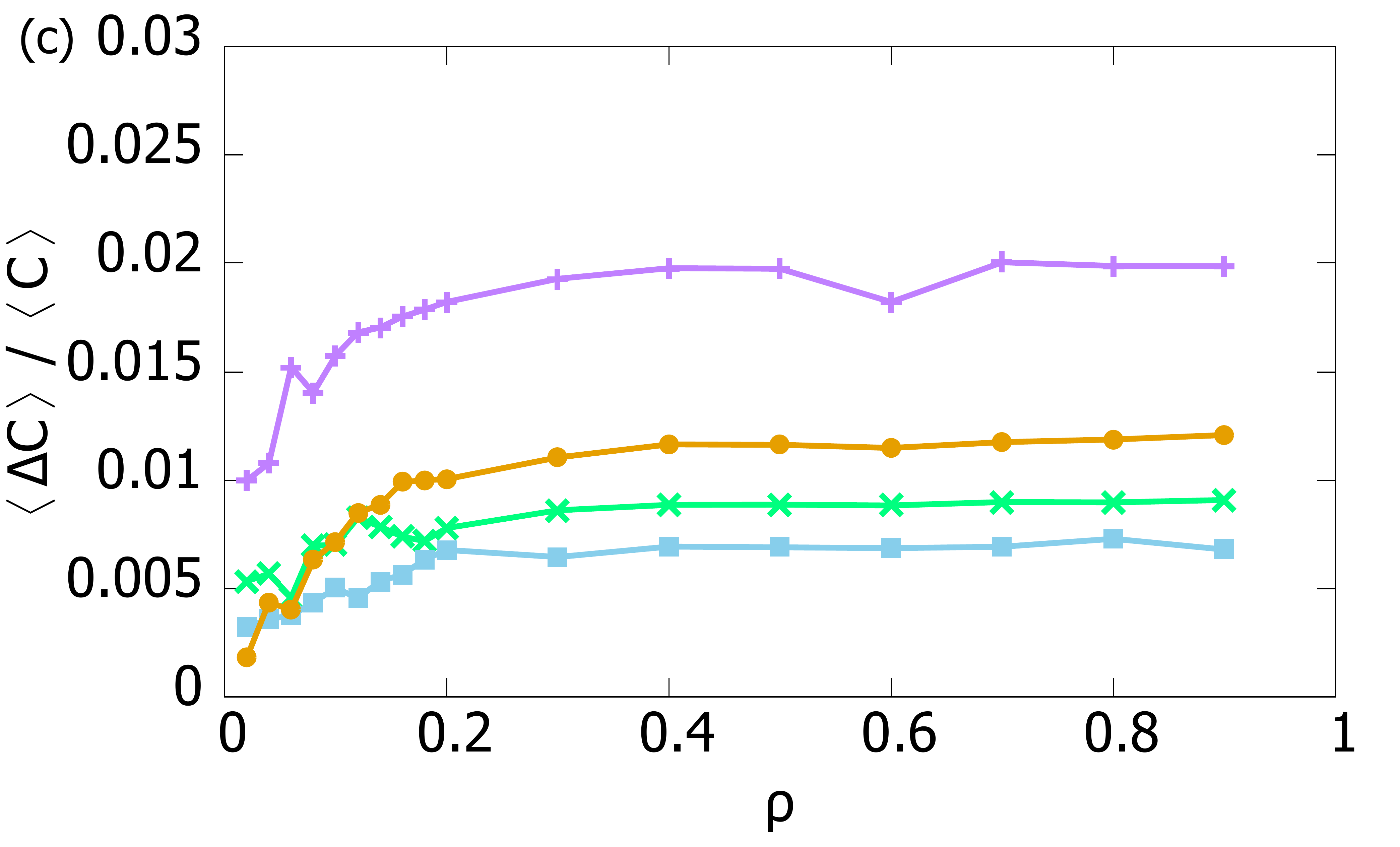}
\caption{
The percentage change in (a) traveling path, (b) traveling distance, and (c) traveling cost as a function of density of vehicles $\rho$ for various network topologies including random regular graphs with different connectivity $k$ and square lattices.
}
\label{fig_diffk}
\end{figure}

Next, we examine the dependence of the optimally diverted traffic on network topology. We first compare the results on random regular graphs with average node degree $k=3, 4$ and $5$ and a single broken link, i.e. $B=1$. As we can see in Fig. \ref{fig_diffk}(a) and (b), both the percentage change of the traveling path and distance are the highest for random graphs with $k=3$, since the average traffic flows on individual links in this case are usually larger than those in $k=4$ or $5$. 

\begin{figure}
\includegraphics[ width=0.9\linewidth] {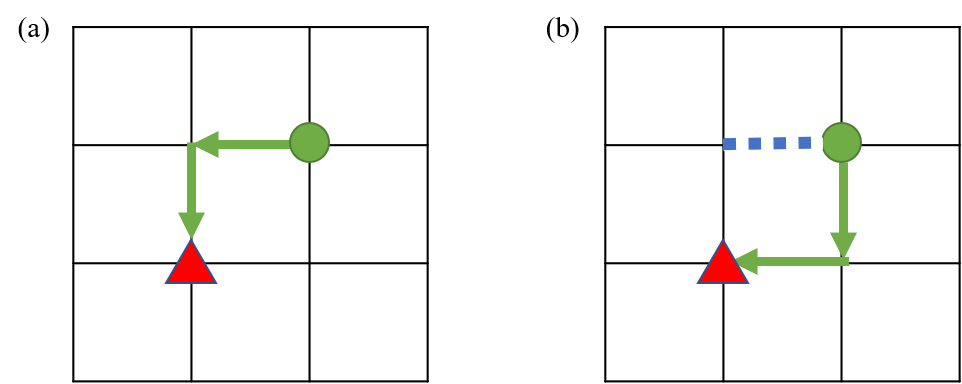}
\caption{
An example of square lattices (a) without broken links and (b) with a single broken link. The vehicle (green circle) re-routes to the destination (red triangle) when a link (blue dotted link) is in its original route is broken. We can see that it is easy for vehicles in square lattices to have alternative paths of the same distance from the destination.
}
\label{fig_exampleSqLattice}
\end{figure}

We then compare the results of random regular graphs with $k=4$ to those from square lattices, since both graphs have four neighbors for each node. As we can see from Fig. \ref{fig_diffk}(a) and (b), $\catpp$ and $\catdp$ from random regular graphs with $k=4$ are higher than those from square lattices. As shown in Fig. \ref{fig_exampleSqLattice}, in square lattices, there is always an alternative route with the same distance from the destination when there is only a single broken link in the network. However, in random graphs, there is not necessarily an alternative path with the same distance. In other words, blocked roads have a larger impact on random regular graphs than on square lattices. Despite such differences, both topologies have similar quantitative behavior in terms of $\catpp$ and $\catdp$, i.e. at a small density $\rho$, both $\catpp$ and $\catdp$ are small; when $\rho$ increases, both $\catpp$ and $\catdp$ increases and then decreases.

Finally, we show the percentage change in traveling cost $\catcp$ in different network topologies in Fig. \ref{fig_diffk}(c). We see that $\catcp$ increases gradually from small $\rho$ and then becomes steady as $\rho$ further increases,  and is highest in random regular graphs with $k=3$, followed by square lattices, and random regular graphs with $k=4$ and $5$. We can also see that the values of $\catcp$ are similar  in random regular graphs with $k=4$ and $5$, implying that the impact of the broken link decreases as connectivity increases, since there are more alternative routes.

\subsection{The benefits brought by coordinated traffic diversion}
\label{sec_compare}
\subsubsection{Benefits on random regular graphs}
After revealing how the optimally diverted traffic depends on various factors, we further examine the benefit from the cases of coordinating diverted traffic, compared to those without coordination. Here, we compare the values of $\catpp$, $\catdp$ and $\catcp$ in cases with $\gamma=1$ in \req{eq_H_broken}, in which vehicles travel with the shortest paths to the destination in an un-coordinated manner, with cases of $\gamma=2$, in which diverted routes are coordinated to suppress congestion \cite{Yeung13717}. 

In the inset of Fig. \ref{fig_changeOfTravelingDistanceAndPath}(a), we see that $\catpp$ in the cases with $\gamma=1$ is larger than that in the cases with $\gamma=2$, suggesting that vehicles diverted in an un-coordinated manner changes their path more drastically than those with coordination. On the contrary, since vehicles in the cases with $\gamma=1$ travel via the shortest paths, as shown in the inset of \ref{fig_changeOfTravelingDistanceAndPath}(b), their $\catdp$ is smaller than that with $\gamma=2$. Interestingly, when $B$ is small, $\catdp$ in the cases with $\gamma=1$ is negative, implying that the vehicles can travel with paths with a shorter distance compared to the initially optimized configuration, as in the example in Fig. \ref{fig_caseTwo}. However, when $B$ increases, more links are removed from the network and vehicles can no longer travel on paths shorter than their original routes, such that $\catdp$ ultimately becomes positive. 

Nevertheless, a shorter traveling distance does not imply that vehicles can travel with a smaller cost. Here, we still consider $\frac{1}{M}\sum_{(ij)}|I_{ij}|^2$ to be the traveling cost even with $\gamma=1$ in optimization. As shown in inset of  Fig. \ref{fig_changeOfTravelingDistanceAndPath}(c),  $\catcp$ in the cases with $\gamma=2$ is always less than that in the cases with $\gamma=1$. It implies that coordinated traffic diversion suppresses the increase in traveling cost, for instance, suppressing traffic congestion, compared to cases of traffic diversion in an un-coordinated manner. As we can see in the inset of Fig. \ref{fig_changeOfTravelingDistanceAndPath}(c), coordinated traffic diversion can save up to $13\%$ of traveling cost when the number of broken links is small, to $1\%$ of the cost when there are many broken links.

\subsubsection{Benefits on the highway network in England}

\begin{figure*}
\includegraphics[ width=0.32\linewidth] {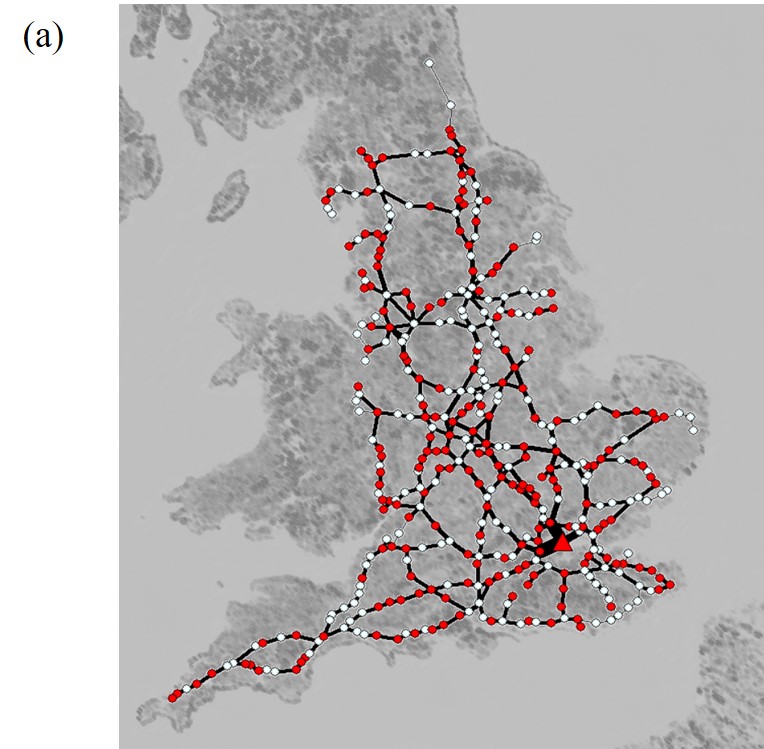}
\includegraphics[ width=0.32\linewidth] {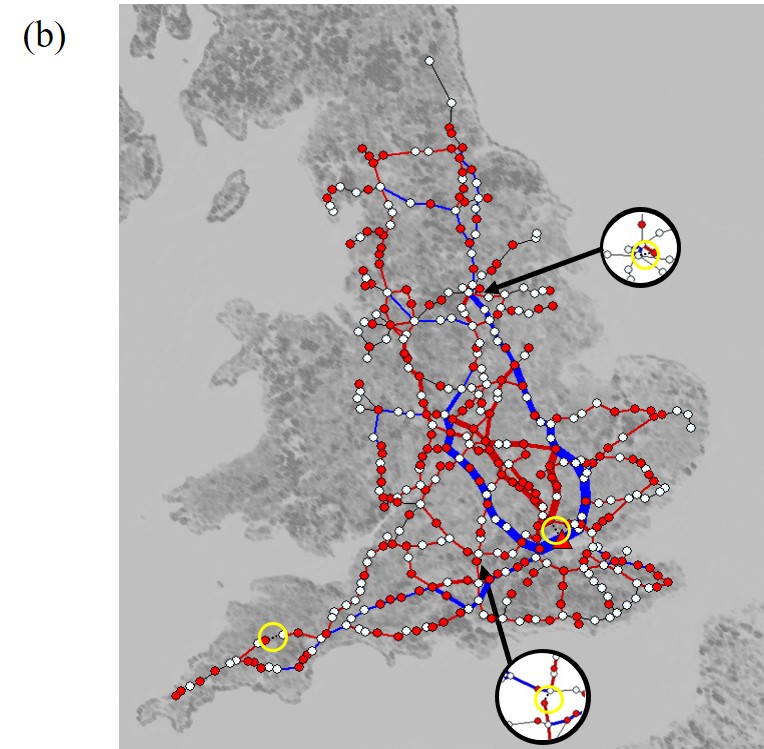}
\includegraphics[ width=0.32\linewidth] {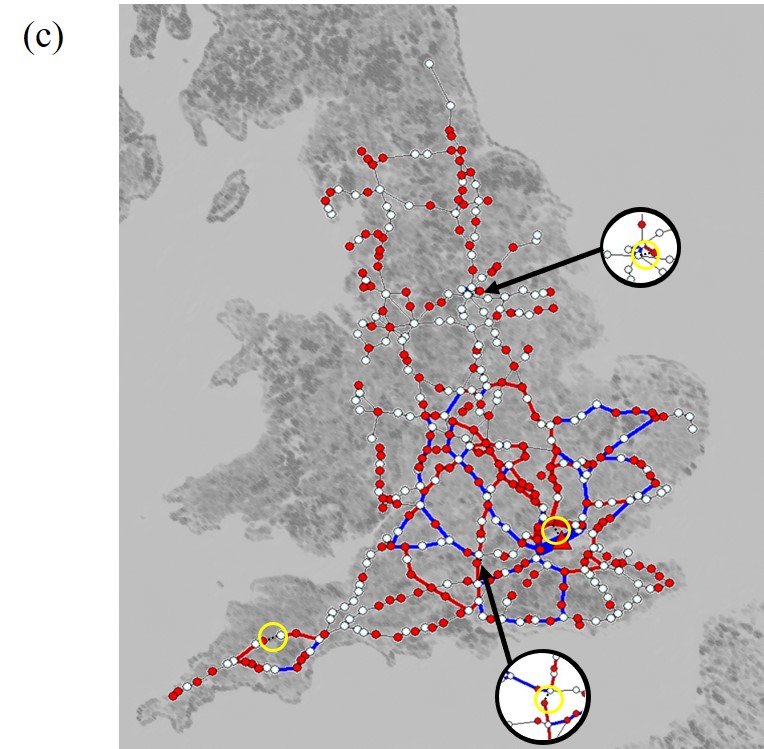}
\caption{
(a) The optimized path configuration for $M=200$ vehicles traveling from different origins in England to London on the highway with $N=395$ nodes. The corresponding path configuration optimized by \req{eq_tE_recur} with (b) $\gamma=1$ and (c) $2$ after four links are broken, i.e. $B=4$. Filled (red) circles correspond to the origins of vehicles, and the triangle represents their common destination in London; the dotted links in (b) and (c) represents the broken links. The thickness of links in (a) represents the magnitude of traffic flow, and the thickness of links in (b) and (c) represents  the magnitude of changes of traffic flow after diversion; blue, red and black links in (b) and (c) represent an increase, decrease or no change in traffic. The route configurations with $\gamma=2$ on average saves as much as $66\%$ of traveling cost compared those with $\gamma=1$ over 100 realizations. Map background is obtained from \copyright 2021 Google.
}
\label{fig_real}
\end{figure*}

We further examine the benefit of coordinated traffic diversion on the highway network in England. Here, we convert the polyline data of the region of the England highway network to a network of $N=395$ nodes \cite{PhysRevE.103.022306}. Figure \ref{fig_real}(a) shows the initial optimized traffic configuration from $\rho N=200$ vehicles traveling from different origins in England to London. Figure \ref{fig_real}(b) and (c) show the un-coordinated and the coordinated changes in traffic flow, i.e. with $\gamma=1$ and $2$ respectively, after four links are blocked in the highway network.  As we can see, the magnitude of changes (as represented by the thickness of links) in (c) is smaller than that in (b), similar to our observation in the inset of Fig. \ref{fig_changeOfTravelingDistanceAndPath}(a) on random regular graphs. We also computed the traveling cost in (b) and (c) and found the case with $\gamma=2$ on average saves as much as $66\%$ of traveling cost compared with the case with $\gamma=1$ over 100 realizations, suggesting a large benefit brought by the coordination of traffic diversion in real networks.

\subsection{Computational cost}
\label{sec_computation}

In this section, we will examine the computational cost of the proposed traffic diversion algorithm Eq.~\ref{eq_tE_recur}. As we have discussed in Sec.~\ref{sec_opt_diversion}, instead of our newly proposed algorithm \req{eq_tE_recur}, the original traffic optimization algorithm \req{eq_prl_recur} can be used to obtain the new traffic configuration in the network with broken links. We first remark that the range of the argument $\dI_{ij}$ in Eq.~(\ref{eq_tE_recur}) is $[-\sfb, \sfb]$, which is smaller than $[-M, M]$ of $I_{ij}$ in Eq. (\ref{eq_prl_recur}). To quantitatively reveal the computational advantage of the newly proposed \req{eq_tE_recur} over \req{eq_prl_recur}, we compare the computational cost in using (A) Eq.~(\ref{eq_prl_recur}), (B) Eq.~(\ref{eq_tE_recur}) with $\dI$ bounded by $[-M, M]$, and (C) Eq.~(\ref{eq_tE_recur}) with $\dI$ bounded by $[-\sfb, \sfb]$. Both (B) and (C) use the traffic configuration $I_{ij}^*$ from the original network as the initial condition for \req{eq_tE_recur}. 

\begin{figure*}
\includegraphics[ width=0.3\linewidth] {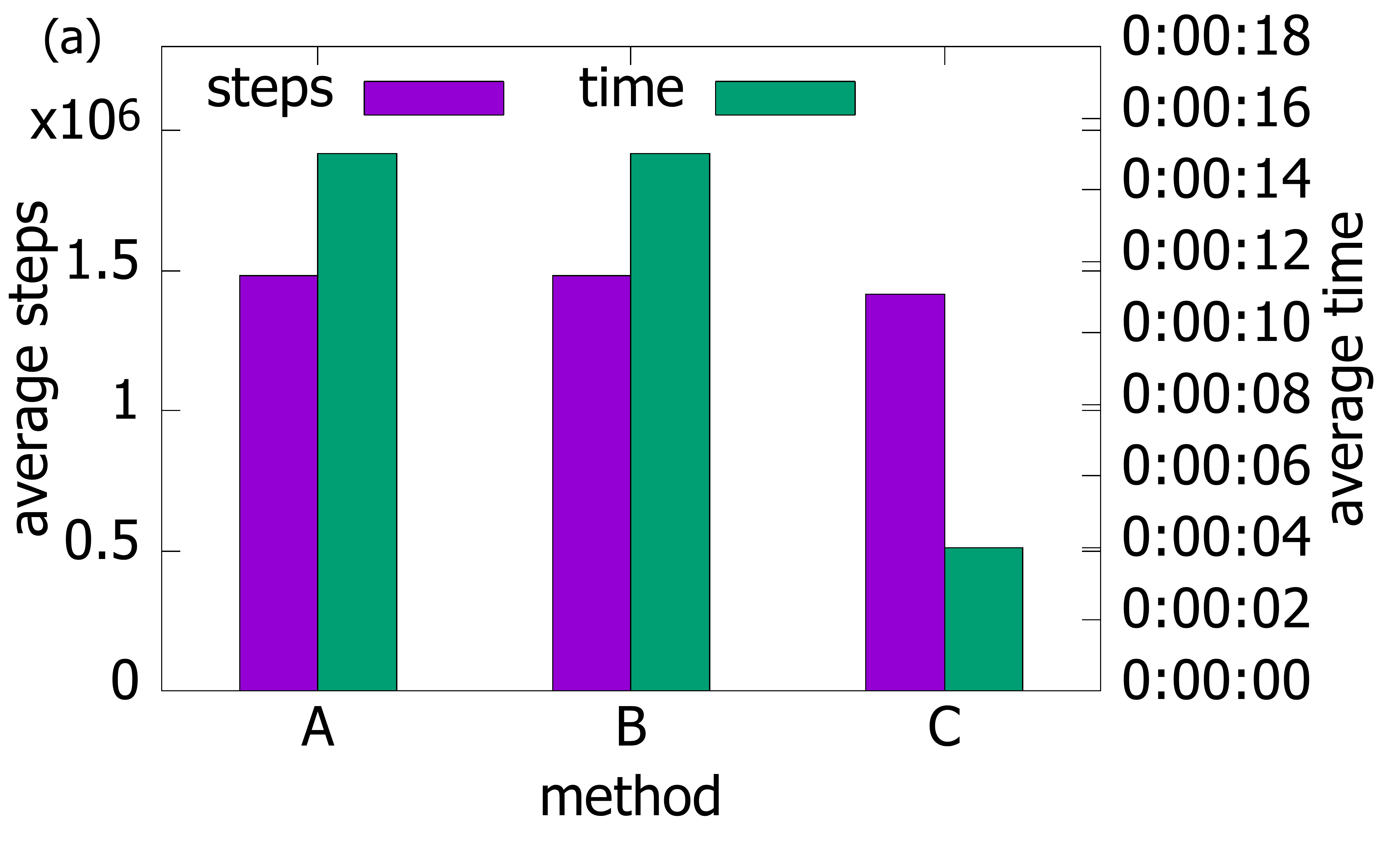}
\includegraphics[ width=0.3\linewidth] {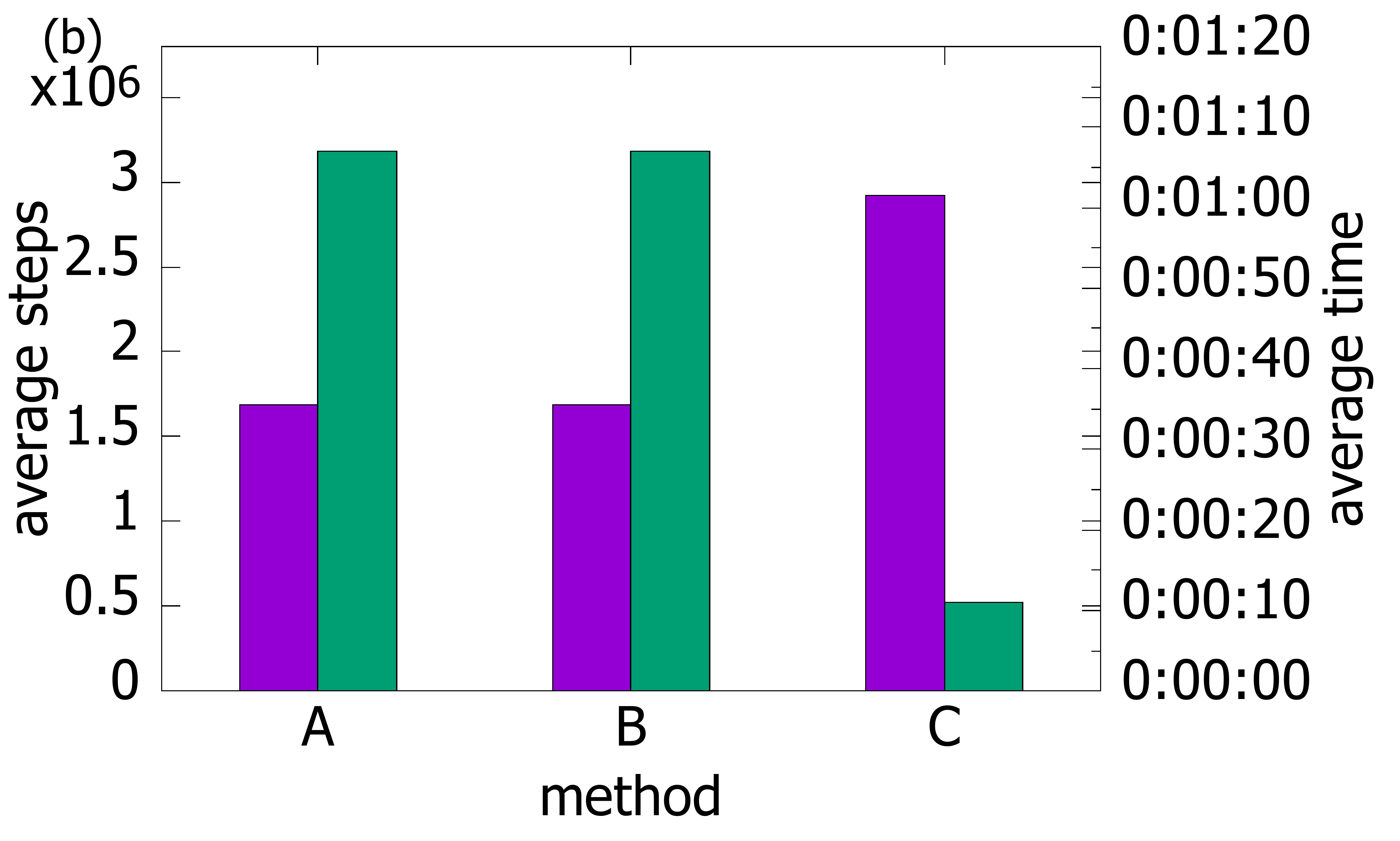}
\includegraphics[ width=0.3\linewidth] {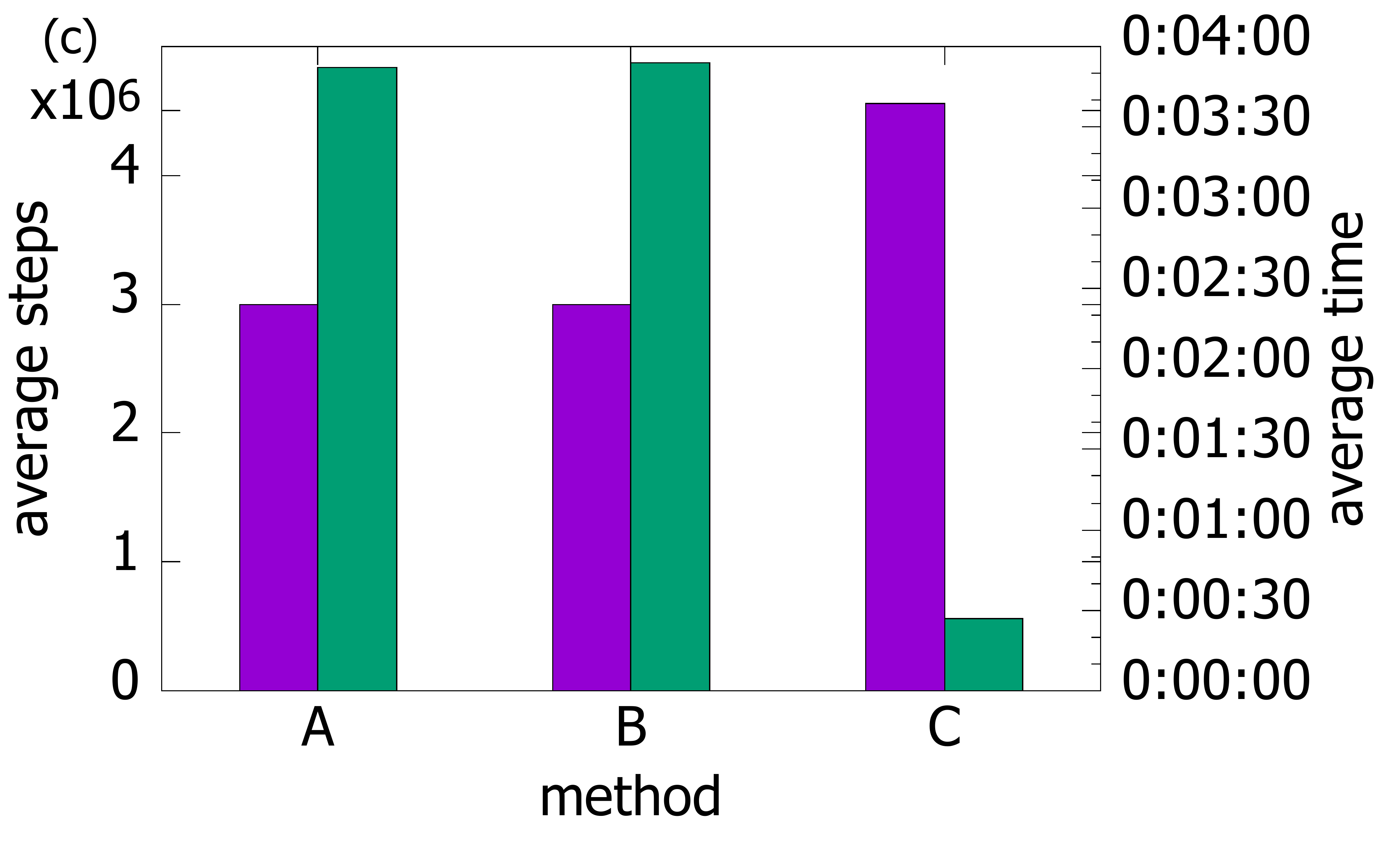}
\caption{
The average number of simulation steps for convergence and the computational time of the three different methods over 1000 realizations, with $B=1$ and (a) $\rho=0.1$, (b) $\rho=0.5$, and (c) $\rho=0.9$. We can see that the average computational time of method (C) is the shortest.
}
\label{fig_diffMethod}
\end{figure*}

As shown in \fig{fig_diffMethod}, all three methods converge after a similar number of simulation time steps at $\rho=0.1$, and (C) uses more steps at $\rho=0.5$ and $\rho=0.9$. However, the number of time steps do not fully reflect the computational time. We thus measure the computational time of the three methods, for simulations conducted on a single core of Intel(R) Xeon(R) CPU E5-2630 v4 @ 2.20GHz. As we can see in \fig{fig_diffMethod}, the computational time of (C) is less than the computational time of (A) and (B) and this difference increases with $\rho$. These results show that even though (C) needs more steps to converge, with the reduced range of argument $\dI$, its overall computational time can be reduced, which is a large advantage of using our proposed algorithm, i.e. \req{eq_tE_recur}.

\subsection{Distribution of traffic flow and discrepancies between analytical and simulation results}
\label{sec_discrepancy}

\begin{figure*}
\includegraphics[ width=0.4\linewidth] {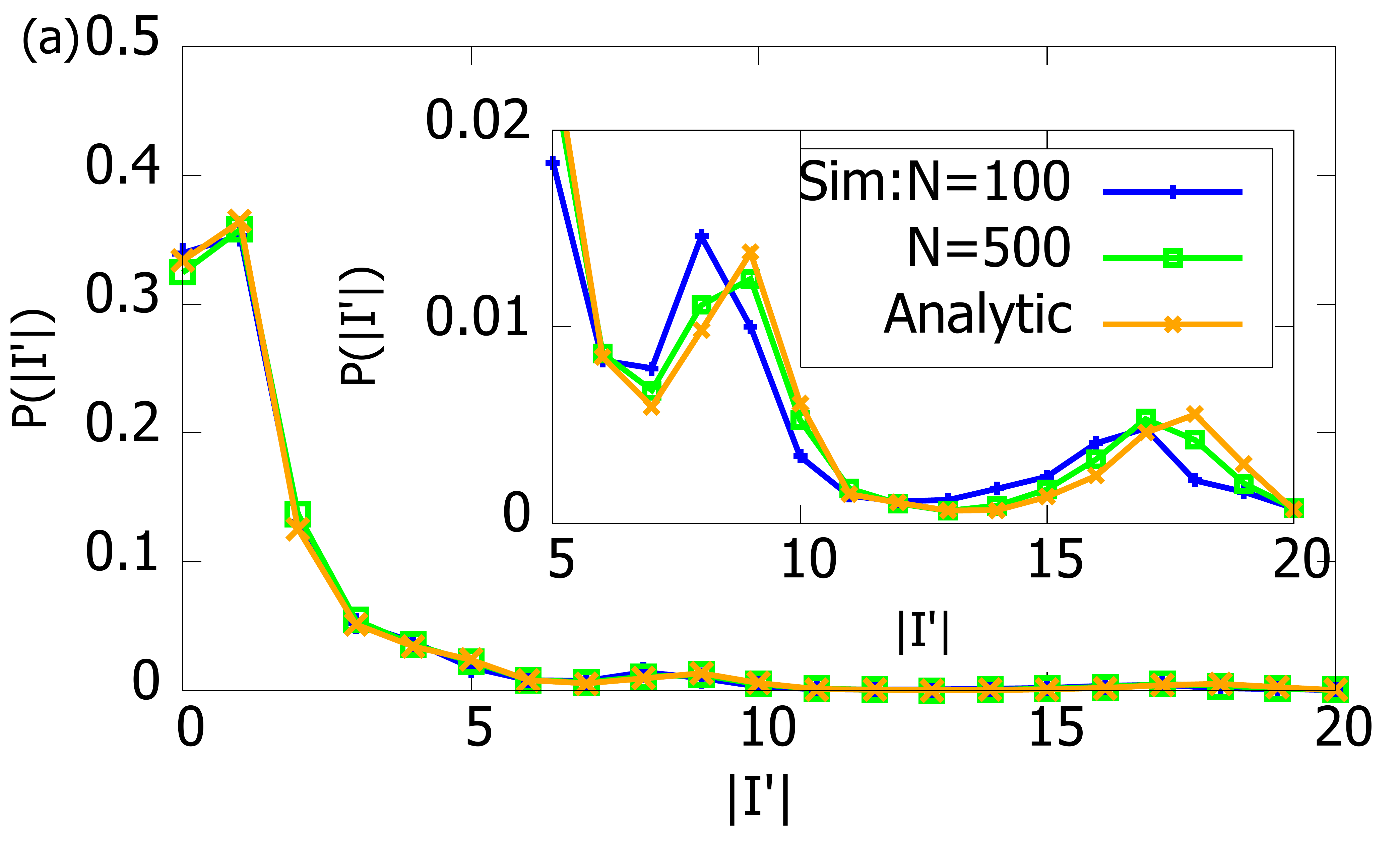}
\includegraphics[ width=0.4\linewidth] {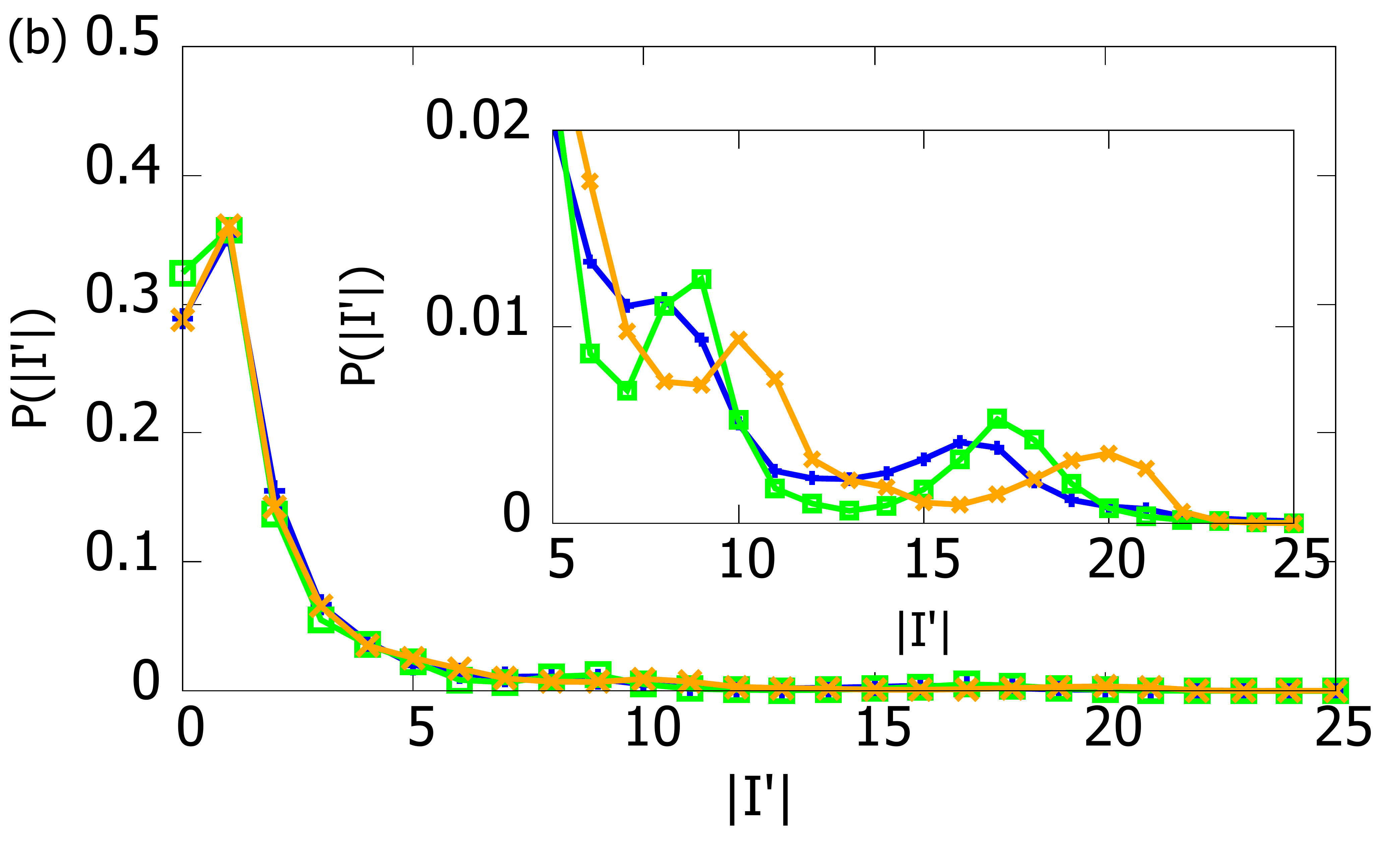}
\caption{
The analytical and simulation results of the distribution of flow $|I'|$ in the network with broken links in cases with (a) $\rho=0.5$ and $B=1$, and (b) $\rho=0.5$ and $B=10$, for random regular graphs with $N=100$ nodes, $k=3$ neighbors and one common destination, compared to the cases with $N=500$ nodes, $k=3$ neighbors and five common destinations. We can see that the agreement between the analytical results and simulation results with $N=500$  is better than that with $N=100$.
}
\label{fig_IDistribution}
\end{figure*}

As shown in \fig{fig_changeOfTravelingDistanceAndPath}, the analytical results of the average percentage change in traveling path, distance and cost, i.e. $\catpp$, $\catdp$ and $\catpp$, capture the trend of the simulation results but there are discrepancies. To understand the discrepancies, in addition to these macroscopic quantities, we look at microscopic quantities such as the distribution of diverted traffic flow $I'$, i.e. $P(I')$. In Fig. \ref{fig_IDistribution}, we  see that the analytical $P(I')$ with $B=1, 10$ agree well with those from simulations ; discrepancies are observed only when we expand the vertical axis to examine the range of $|I'|\ge 5$ as shown in the insets of Fig. \ref{fig_IDistribution}.

\begin{figure*}
\includegraphics[ width=0.4\linewidth] {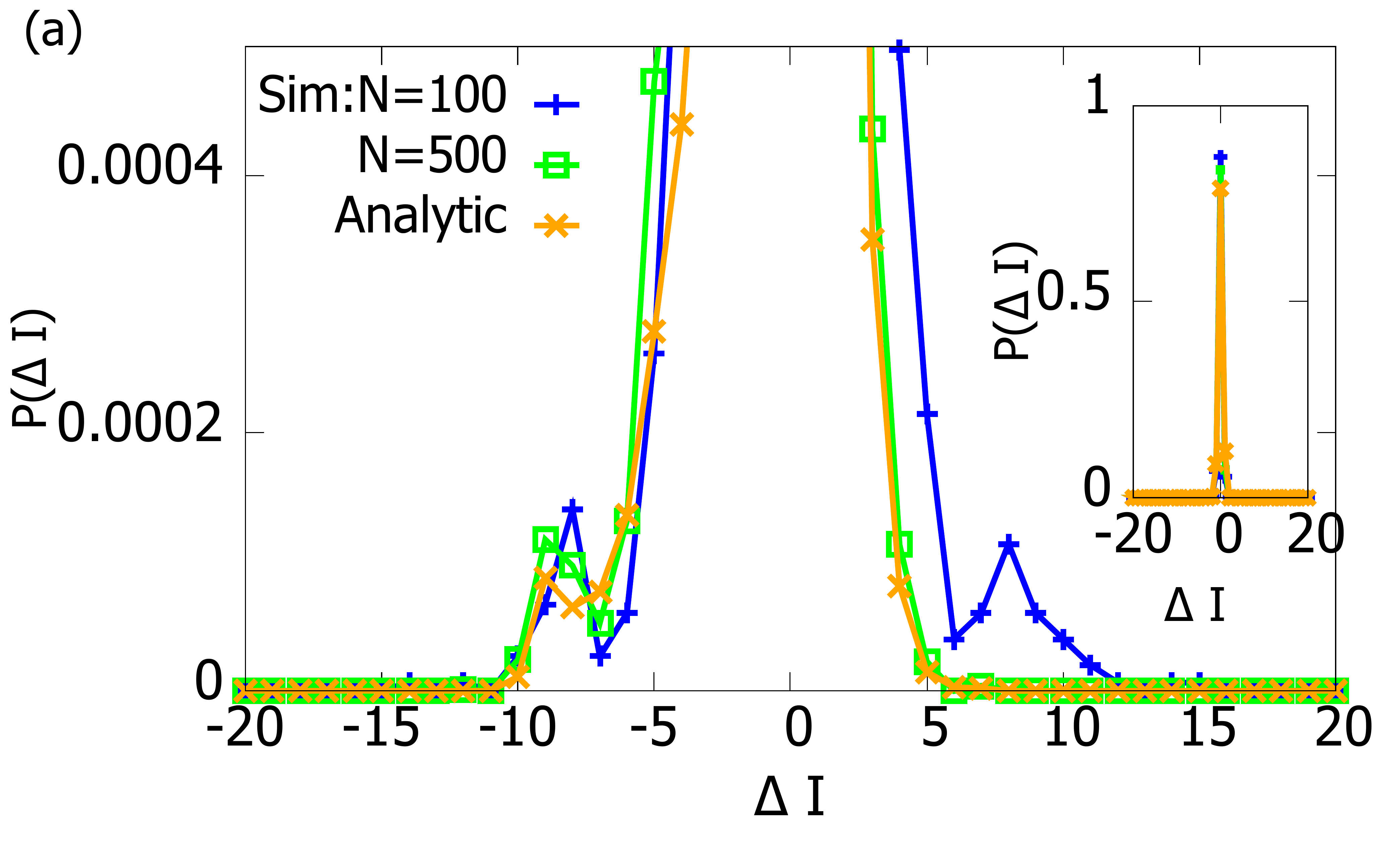}
\includegraphics[ width=0.4\linewidth] {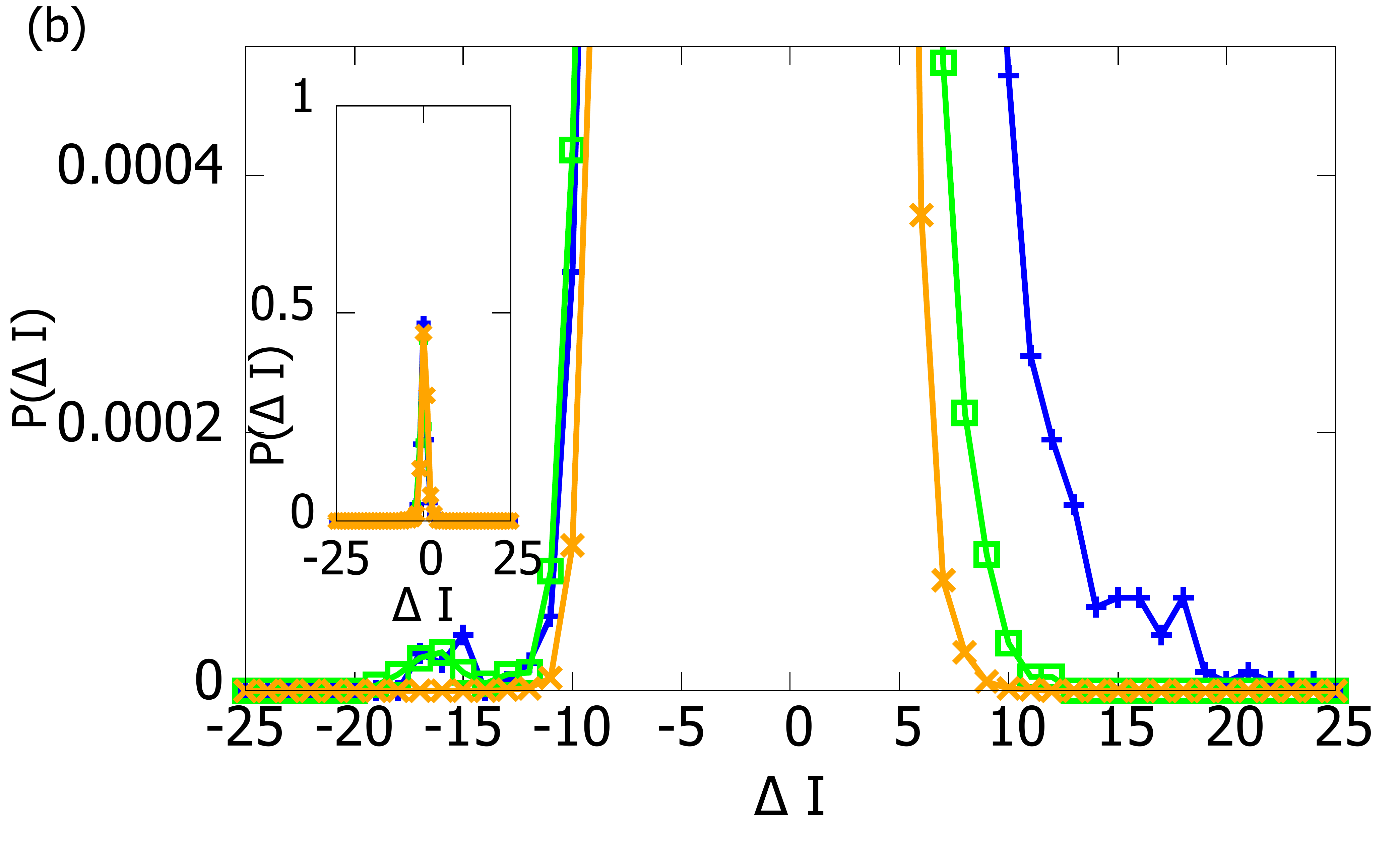}
\caption{
The analytical and simulation results of the distribution of change of traffic flow $\dI$ in  the network with broken links in the cases with (a) $\rho=0.5$ and $B=1$, and (b) $\rho=0.5$ and $B=10$,  for random regular graphs with $N=100$ nodes, $k=3$ neighbors and one common destination, compared to the cases with $N=500$ nodes, $k=3$ neighbors and five common destinations, and analytical solutions. Similar to \fig{fig_IDistribution}, the agreement between the analytical results and simulation results improves with $N=500$ compared to that with $N=100$.
}
\label{fig_DeltaIDistribution}
\end{figure*}

Since our analytical results are derived in the thermodynamics limit, we suggest that finite-size effect in simulations as the origin of these observed discrepancies. As we can see in the insets of \fig{fig_IDistribution}, the simulation results with $N=500$ agree better with the analytical results compared to that with $N=100$. We remark that the ratio of common destinations, vehicles and broken links are preserved when system size increases in simulations. The peaks in $P(I')$ are wider and shifted slightly to larger values of $I'$ when $N$ increases, since vehicles have a larger choice of destinations, resulting in less similar configuration of flow among each destination, which is an evident of finite-size effect. As shown in Fig. \ref{fig_DeltaIDistribution}, the agreement between simulations and analytical distributions of changes in flow on a link, i.e. $P(\dI)$, also improves with the system size.

\section{Conclusion}
In this paper, we introduced a model of transportation networks in which vehicle routes are initially optimized before a set of links fail to function.  We then applied statistical physics to derive both analytical solutions and an optimization algorithm to optimally divert and coordinate the traffic due to the broken links. Our results show that coordinated traffic diversion after partial network failure suppresses the increase in traveling cost compared to the case without coordination. Since vehicles travel via the shortest path to the destination when they are un-coordinated, when the number of broken links is small, the average traveling distance may decrease compared to the initially optimized configuration; nevertheless, traveling cost increases and can be much higher than the case with coordinated traffic diversion. The magnitude of route changes is smaller in the coordinated case compared to the un-coordinated one. Moreover, the closer the broken links to the common destination, the higher the impact on the system. When the number of broken links increases, the average change in traveling path, distance and cost increases. We also see that when the connectivity of nodes increases, the impact of the broken links decreases since the number of alternative routes increases.

In summary, our results show that coordinated traffic diversion is beneficial to transportation networks with broken links, which saves as much as $13\%$ of the  traveling cost in random regular graphs compared to the case without coordination. By testing our algorithm on the highway network in England, we obtain qualitative similar behavior and our derived algorithm saves up to $66\%$ of the traveling cost. These results shed light on the importance of coordination in diverting traffic after partial network failure.

\begin{acknowledgments}
This work is supported by the Research Grants Council of the Hong Kong Special Administrative Region, China (Projects No. EdUHK ECS 28300215, No. GRF 18304316, No. GRF 18301217, and No. GRF 18301119); the EdUHK FLASS Dean's Research Fund Grants No. IRS12 2019 04418 and No. ROP14 2019 04396; and EdUHK RDO Internal Research Grants No. RG67 2018-2019R R4015 and No. RG31 2020-2021R R4152
\end{acknowledgments}


\bibliographystyle{h-physrev4}


\end{document}